\documentclass[10pt, journal]{IEEEtran}

\usepackage{graphicx, enumerate}
\usepackage[OT1]{fontenc}
\usepackage{cite, array, bbm, svg}
\usepackage{dsfont, epstopdf}
\usepackage{mathrsfs, subfigure, color}
\usepackage{amssymb, multirow}
\usepackage[cmex10]{amsmath}
\usepackage{subeqnarray, tabularx}
\usepackage{cases, stmaryrd}
\usepackage{algpseudocode}
\usepackage[ruled, vlined, linesnumbered]{algorithm2e}
\usepackage[utf8]{inputenc}
\usepackage{threeparttable}
\usepackage{stackengine}
\usepackage{verbatim}
\usepackage[bbgreekl]{mathbbol}

\SetKwInput{KwInit}{Initialize}
\usepackage[mathscr]{eucal}

\newcommand{\overbar}[1]{\mkern 1.5mu\overline{\mkern-1.5mu#1\mkern-1.5mu}\mkern 1.5mu}

\DeclareSymbolFontAlphabet{\mathbb}{AMSb}
\DeclareSymbolFontAlphabet{\mathbbb}{bbold}

\DeclareFontFamily{U}{FdSymbolC}{}
\DeclareFontShape{U}{FdSymbolC}{m}{n}{<-> s * FdSymbolC-Book}{}
\DeclareSymbolFont{fdarrows}{U}{FdSymbolC}{m}{n}
\DeclareMathSymbol{\vDdash}{\mathrel}{fdarrows}{254}

\DeclareFontFamily{U}{FdSymbolD}{}
\DeclareFontShape{U}{FdSymbolD}{m}{n}{<-> s * FdSymbolD-Book}{}
\DeclareSymbolFont{fdnarrows}{U}{FdSymbolD}{m}{n}
\DeclareMathSymbol{\nvDdash}{\mathrel}{fdnarrows}{254}
\makeatother

\newtheorem{theorem}{Theorem}

\newtheorem{lemma}{Lemma}
\newtheorem{proposition}{Proposition}
\newtheorem{definition}{Definition}
\newtheorem{remark}{Remark}
\newtheorem{example}{Example}

\DeclareMathOperator{\Id}{\textrm{Id}}

\DeclareMathOperator{\cstr}{\textrm{c}}

\DeclareMathOperator{\enab}{\mathsf{enab}}

\DeclareMathOperator{\reach}{\mathsf{Reach}}

\begin{document}

\title{\LARGE \bf Zonotope-based Symbolic Controller Synthesis \\ for Linear Temporal Logic Specifications
\thanks{This work was supported by the Fundamental Research Funds for the Central Universities under Grant DUT22RT(3)090, the National Natural Science Foundation of China under Grant 08120003, the H2020 ERC Consolidator Grants L2C (Grant 864017) and LEAFHOUND (Grant 864720), the CHIST-ERA 2018 project DRUID-NET, the Walloon Region and the Innoviris Foundation, the Swedish Research Council (VR) and the Knut och Alice Wallenberg Foundation (KAW).}
}

\author{Wei~Ren, \IEEEmembership{Member, IEEE}, Rapha\"el M. Jungers, \IEEEmembership{Senior Member, IEEE}, and Dimos V. Dimarogonas, \IEEEmembership{Fellow, IEEE}
\thanks{W. Ren is with Key Laboratory of Intelligent Control and Optimization for Industrial Equipment of Ministry of Education, Dalian University of Technology, Dalian 116024, China. R. M. Jungers is with ICTEAM institute, UCLouvain, 1348 Louvain-la-Neuve, Belgium. D. V. Dimarogonas is with Division of Decision and Control Systems, EECS, KTH Royal Institute of Technology, SE-10044, Stockholm, Sweden.
Email: \texttt{\small wei.ren@dlut.edu.cn, raphael.jungers@uclouvain.be, dimos@kth.se}.}
}

\maketitle

\begin{abstract}
This paper studies the controller synthesis problem for nonlinear control systems under linear temporal logic (LTL) specifications using zonotope techniques. A local-to-global control strategy is proposed for the desired specification expressed as an LTL formula. First, a novel approach is developed to divide the state space into finite zonotopes and constrained zonotopes, which are called \emph{cells} and allowed to intersect with the neighbor cells. Second, from the intersection relation, a graph among all cells is generated to verify the realization of the accepting path for the LTL formula. The realization verification determines if there is a need for the control design, and also results in finite local LTL formulas. Third, once the accepting path is realized, a novel abstraction-based method is derived for the controller design. In particular, we only focus on the cells from the realization verification and approximate each cell thanks to properties of zonotopes. Based on local symbolic models and local LTL formulas, an iterative synthesis algorithm is proposed to design all local abstract controllers, whose existence and combination establish the global controller for the LTL formula. Finally, the proposed framework is illustrated via a path planning problem of mobile robots.
\end{abstract}

\begin{IEEEkeywords}
Controller synthesis, nonlinear systems, symbolic control, linear temporal logic, zonotope-based covering.
\end{IEEEkeywords}

\section{Introduction}
\label{sec-intro}

Due to the resemblance to natural language and the existence of off-the-shelf algorithms for model checking \cite{Baier2008principles}, linear temporal logic (LTL) is an expressive language to specify high-level tasks for dynamical systems. Because of the complexities of tasks and dynamical systems, control synthesis problems usually cannot be solved directly. For this purpose, many approaches have been proposed, such as sampling/optimization-based approaches \cite{Wolff2014optimization}, game-based approaches \cite{Bloem2012synthesis, Clarke2018handbook}, and automata/abstraction-based approaches \cite{Wongpiromsarn2011tulip}. Among all these approaches, the abstraction-based approach allows us to solve control synthesis problems efficiently via techniques from supervisory control \cite{Antsaklis1993hybrid} or game theory \cite{Bloem2012synthesis}. Symbolic abstractions are a dynamical systems description where each symbolic state is a collection or aggregate of system states \cite{Tabuada2009verification}. With an equivalence relation between the dynamical system and its symbolic abstraction, the controller is synthesized to be correct by design \cite{Girard2012controller}, and the formal verification may be unnecessary.

In the abstraction-based control approach, the bottom-up strategy is standard and has two steps: the abstraction construction to discretize dynamical systems, and the backward searching to find abstract controllers. Once an equivalence relation between the dynamical system and its abstraction is established, the abstract controller can be refined as a hybrid controller for the dynamical system \cite{Tabuada2009verification}. Since the backward searching can be done by existing algorithmic tools \cite{Antsaklis1993hybrid, Bloem2012synthesis}, the key of the bottom-up strategy is the abstraction construction to ensure certain equivalence relations. Many classes of dynamical systems admitting symbolic abstractions have been identified, including nonlinear systems \cite{Pola2008approximately, Zamani2012symbolic, Ren2019logarithmic}, time-delay systems \cite{Pola2010symbolic, Ren2020symbolic}, switched systems \cite{Girard2016safety}, networked control systems\cite{Zamani2017symbolic} and stochastic systems \cite{Zamani2015symbolic}. In terms of equivalence relations, (bi-)simulation relation \cite{Tabuada2009verification}, feedback refinement relation \cite{Reissig2017feedback} and their variants/extensions \cite{Pola2008approximately, Kim2017symbolic} have been widely studied. Recently many toolboxes \cite{Mazo2010pessoa, Mouelhi2013cosyma, Rungger2016scots, Wongpiromsarn2011tulip} have been developed for high-level control problems.

Despite these works via the abstraction-based control approach, the control synthesis for dynamical systems under LTL specifications is still challenging from the following two perspectives. Regarding dynamical systems and LTL formulas, general LTL formulas or GR(1) formulas \cite{Bloem2012synthesis} are only studied for dynamical systems with simple structures \cite{Tumova2010symbolic, Wongpiromsarn2011tulip}, or only simple/specific LTL formulas like safety and reachability specifications are addressed for complex dynamical systems \cite{Mazo2010pessoa, Mouelhi2013cosyma, Rungger2016scots}. A direct question is: \emph{how do we deal with the controller synthesis problem for complex dynamical systems under general LTL formulas}? Regarding the control synthesis, the design strategies are usually derived in a global way that the state space of dynamical systems needs to be considered as a whole \cite{Tabuada2009verification, Reissig2017feedback}. In this respect, although different equivalence relations are proposed, the abstraction construction and the controller synthesis are still time-consuming and computationally complex. Hence, how to address the controller synthesis problem in a local way needs further study. 

Motivated by the above discussion, in this paper we propose a zonotope-based top-down approach to deal with the control synthesis problem for dynamical systems under LTL specifications. In particular, we propose a zonotope-based approach to verify the realization of LTL formulas, and then combine the realization verification and the abstraction-based control design to develop a local-to-global control strategy. To this end, our first contribution is the zonotope-based approach for the realization of LTL formulas. Different from many works \cite{Scott2016constrained, Girard2005reachability, Mitchell2019invariant} where zonotopes are used to represent or estimate reachable sets, we cover the state space via finite zonotopes and constrained zonotopes, which are called \emph{cells} and allowed to intersect with neighbor cells. Based on the intersection relation among all cells, a graph is constructed and further applied to verify the realization of the LTL formula by searching an admissible path to realize the accepting path of the LTL formula. From the graph-based method, the realization verification determines a sequence of finite cells, which  not only shows the necessity of the controller design, but also prepares for the controller synthesis afterwards.

Due to the zonotope-based approach and the realization verification, a direct outcome is a formal decomposition of the satisfaction of the global LTL formula into the satisfaction of finite local ones. Hence, the control synthesis problem is decomposed into finite local ones, which can be solved iteratively. Another outcome is a novel and local abstraction construction approach. In particular, each cell is approximated locally via its own properties, and thus the approximations of different cells are not necessarily the same. Furthermore, different local symbolic models are constructed such that different equivalence relations can be satisfied. The proposed approach is local such that different equivalence relations can be satisfied locally, which potentially reduces the conservatism caused by one equivalence relation and the computational complexity from the global approximation.

With the decomposition of the control synthesis problem and the construction of local symbolic models, a local-to-global control strategy is developed to design all local abstract controllers to ensure the satisfaction of the LTL formula, which is the second contribution of this paper. For each cell, its intersection with the previous cell is the initial region, whereas its intersection with the next cell is the target region. Hence, each cell has its local LTL formula, which is combined with the local symbolic model to design the local controller. From the local LTL formula of each cell, the local controller design strategy is implemented repeatedly, which is formulated into an iterative synthesis algorithm. In addition, using the derived equivalence relations, we further discuss the refinement of local abstract controllers and the relation between local and global controllers, which validates the satisfaction of the LTL formula via all designed local abstract controllers.

A preliminary version of this work has been presented in \cite{Ren2021zonotope}, and is expanded significantly here in several directions. First, since we aim to propose a local-to-global control strategy, here the realization verification is modified and the detailed decomposition is established to show how to decompose the LTL formula into finite local ones. Hence, the robust accepting path and its realization verification are presented to ensure the well-decomposedness of the LTL formula. Second, the controller design is only outlined in \cite{Ren2021zonotope} while presented here in detail. In terms of the abstraction construction, we address both the case that the system is not incrementally stable and the case that the system is incrementally stable, while only the former one is studied in \cite{Ren2021zonotope}. In terms of the controller design, the relation between the global and local control problems is derived, and an explicit local-to-global control strategy is established. Therefore, different equivalence relations can be mixed up in the current setting, which however is not the case in \cite{Ren2021zonotope}. Finally, all formal proofs of the main results are included in this paper to substantiate the correctness and feasibility of the proposed control strategy.

The rest of this paper is organized as follows. Preliminaries are stated in Section \ref{sec-notepre} and the problem formulation is given in Section \ref{sec-problem}. Section \ref{sec-partition} presents the covering mechanism, and Section \ref{sec-verifydiscrete} shows the realization verification. The controller synthesis is derived in Section \ref{sec-abstractionbased}. Simulation results are given in Section \ref{sec-example}. Section \ref{sec-conclusion} concludes with future work.

\section{Related Work}
\label{sec-relatedwork}

The control synthesis problem for LTL specifications has been considered extensively \cite{Belta2007symbolic, Kress2009temporal, Fainekos2009temporal, Wongpiromsarn2011tulip, Gol2013time, Meyer2019hierarchical}. In abstraction-based control, the bottom-up and top-down structures are two common control approaches, both of which are built on the partition of the state space. A coarse partition is computed in a top-down strategy to derive a discrete plan for the controller design in \cite{Kress2009temporal, Meyer2019hierarchical}, whereas a finer partition is used in a bottom-up strategy to construct an appropriate symbolic model such that certain equivalence relation is satisfied \cite{Girard2007approximation, Pola2008approximately, Reissig2017feedback, Kim2017symbolic, Ren2019logarithmic}. In these works, neighbor cells are adjacent but not overlapped. Hence, in order to deal with the controller synthesis problem, the state-space partition needs to be considered globally \cite{Girard2016safety, Reissig2017feedback} or semi-globally \cite{Belta2007symbolic, Meyer2019hierarchical}. In this respect, general LTL specifications are hard to be addressed efficiently and the computational complexity is still huge. In this work, we propose a zonotope-based covering mechanism for the state space, and allow for the intersection among neighbor cells such that local LTL specifications can be derived, which can further be resolved iteratively and efficiently via existing tools. 

In the abstraction construction, the bottom-up strategy requires the symbolic abstraction to be constructed first such that a fully actuated model is derived. In this way, the entire state space is considered integrally and uniformly, while some assumptions like incremental stability properties are needed in \cite{Tabuada2008approximate, Girard2010approximately, Pola2008approximately, Girard2012controller} for the well-constructedness. However, global and uniform discretization of high-dimensional dynamical systems usually results in complexity explosion. On the other hand, the top-down strategy allows to construct symbolic abstractions at different levels of granularity to deal with the planning and control problem sequentially. For instance, a coarse abstraction is constructed first to derive a high-level plan over regions of interest in the workspace given the temporal task \cite{Fainekos2009temporal, Guo2015multi, Meyer2019hierarchical}, which is then combined with the abstraction refinement \cite{Hsu2018lazy, Meyer2019hierarchical} to design a low-level feedback controller to execute the derived high-level plan. The abstraction construction or refinement is still (semi-)global, and only one equivalence relation is allowed in these existing works. Due to the zonotope-based covering and the resulting intersection relation, here we propose a novel graph-based way for high-level planning and realization verification, and then construct the symbolic abstraction for each cell individually and locally, which hence allows different equivalence relations to be combined and further facilitates the controller design.

\section{Notation and Preliminaries}
\label{sec-notepre}

Let $\mathbb{R}=(-\infty, +\infty)$, $\mathbb{R}^{+}=[0, +\infty)$, $\mathbb{N}=\{0, 1, \ldots\}$ and $\mathbb{N}^{+}=\{1, 2, \ldots\}$. $\mathbb{R}^{n}$ is the $n$-dimensional Euclidean space. For $x\in\mathbb{R}^{n}$, $x_{i}$ is the $i$-th element of $x$; $\|x\|$ is the infinity norm of $x$. $\mathbbmss{B}(x, \varepsilon)=\{y\in\mathbb{R}^{n}:\|x-y\|\leq\varepsilon\}$ denotes the closed ball centered at $x\in\mathbb{R}^{n}$ with radius $\varepsilon\in\mathbb{R}^{+}$. For a finite set $\mathbb{A}\subset\mathbb{R}^{n}$, $|\mathbb{A}|$ is the cardinality of $\mathbb{A}$. For $\mathbb{A}, \mathbb{B}\subset\mathbb{R}^{n}$, $\mathbb{A}$ is compact if it is closed and bounded; $\mathbb{B}\setminus\mathbb{A}=\{x: x\in\mathbb{B}, x\notin\mathbb{A}\}$. For a compact set $\mathbb{A}\subset\mathbb{R}^{n}$ and $\varepsilon>0$, $\mathbb{A}^{\circ}$ is the interior of $\mathbb{A}$, and $\partial\mathbb{A}$ is the boundary of $\mathbb{A}$; $\mathbb{A}^{\varepsilon}=\{x\in\mathbb{A}: \mathbbmss{B}(x, \varepsilon)\subseteq\mathbb{A}\}$ is the $\varepsilon$-contraction of $\mathbb{A}$; $\mathbf{E}_{\varepsilon}(\mathbb{A})=\{y\in\mathbb{R}^{n}: \exists x\in\mathbb{A}, \|y-x\|\leq\varepsilon\}$ is the $\varepsilon$-expansion of $\mathbb{A}$. A relation $\mathscr{R}\subseteq\mathbb{A}\times\mathbb{B}$ is identified with the map $\mathscr{R}: \mathbb{A}\rightarrow2^{\mathbb{B}}$ defined by $b\in\mathscr{R}(a)$ if and only if $(a, b)\in\mathscr{R}$. The inverse relation of $\mathscr{R}$ is $\mathscr{R}^{-1}=\{(b, a)\in\mathbb{B}\times\mathbb{A}: (a, b)\in\mathscr{R}\}$. $\Id$ is the identity function. A function $\alpha: \mathbb{R}^{+}\to\mathbb{R}^{+}$ is of class $\mathcal{K}$ if it is continuous, strictly increasing and $\alpha(0)=0$; $\alpha$ is of class $\mathcal{K}_{\infty}$ if $\alpha\in\mathcal{K}$ and it is unbounded. A function $\beta: \mathbb{R}^{+}\times\mathbb{R}^{+}\to\mathbb{R}^{+}$ is of class $\mathcal{KL}$ if $\beta(s, t)\in\mathcal{K}$ for each fixed $t\geq0$ and $\beta(s, t)$ decreases to zero as $t\rightarrow\infty$ for each fixed $s\geq0$.

A set $\mathbf{Z}\subset\mathbb{R}^{n}$ is a \emph{zonotope}, if there exists $(\mathbf{c}, \mathbf{G})\in\mathbb{R}^{n}\times\mathbb{R}^{n\times n_{g}}$ such that $\mathbf{Z}=\{\mathbf{c}+\mathbf{G}\xi: \|\xi\|\leq1\}$, where $\mathbf{c}\in\mathbb{R}^{n}$ is the center and $\mathbf{G}\in\mathbb{R}^{n\times n_{g}}$ is the generator matrix with each column as a generator. A set $\mathbf{Z}^{\cstr}\subset\mathbb{R}^{n}$ is a \emph{constrained zonotope}, if there exists $(\mathbf{c}, \mathbf{G}, \mathbf{A}, \mathbf{b})\in\mathbb{R}^{n}\times\mathbb{R}^{n\times n_{g}}\times\mathbb{R}^{n_{c}\times n_{g}}\times\mathbb{R}^{n_{c}}$ such that $\mathbf{Z}^{\cstr}=\{\mathbf{c}+\mathbf{G}\xi: \|\xi\|\leq1, \mathbf{A}\xi=\mathbf{b}\}$, where $\mathbf{A}\xi=\mathbf{b}$ is the constraint condition. We use the notations $\mathbf{Z}=\{\mathbf{c}, \mathbf{G}\}$ (generator representation or G-representation) for zonotopes, and $\mathbf{Z}^{\cstr}=\{\mathbf{c}, \mathbf{G}, \mathbf{A}, \mathbf{b}\}$ (constrained generator representation or CG-representation) for constrained zonotopes. Given $\mathbf{Z}=\{\mathbf{c}, \mathbf{G}\}$, if $\mathbf{G}$ is diagonal, orthogonal or invertible, then $\mathbf{Z}$ is reduced to a box, a hypercube or a parallelotope, respectively. From \cite{Scott2016constrained}, $\mathbf{Z}^{\cstr}\subset\mathbb{R}^{n}$ is a constrained zonotope if and only if it is a convex polytope.

\begin{lemma}[\cite{Scott2016constrained}]
\label{lem-1}
For every $\mathbf{Z}^{\cstr}=\{\mathbf{c}, \mathbf{G}, \mathbf{A}, \mathbf{b}\}\subset\mathbb{R}^{n}$, $\mathbf{Z}^{\cstr}\neq\varnothing$ if and only if $\min\{\|\xi\|: \mathbf{A}\xi=\mathbf{b}\}\leq1$; $z\in\mathbf{Z}^{\cstr}$ if and only if $\min\{\|\xi\|: \mathbf{G}\xi=z-\mathbf{c}, \mathbf{A}\xi=\mathbf{b}\}\leq1$.
\end{lemma}

\subsection{Transition Systems}
\label{subsec-approbisimu}

\begin{definition}[\cite{Girard2007approximation}]
\label{def-1}
A \emph{transition system} is a sextuple $\mathbf{T}=(X, X_{0}, U, \Delta, Y, H)$, consisting of: (i) a set of states $X$; (ii) a set of initial states $X^{0}\subseteq X$; (iii) a set of inputs $U$; (iv) a transition relation $\Delta\subseteq X\times U\times X$; (v) a set of outputs $Y$; (vi) an output function $H: X\rightarrow Y$. $\mathbf{T}$ is said to be \emph{metric} if the output set $Y$ is equipped with a metric $\mathbf{d}: Y\times Y\rightarrow\mathbb{R}^{+}$, and \emph{symbolic} if the sets $X$ and $U$ are finite or countable.
\end{definition}

The transition $(x, u, x')\in\Delta$ is denoted by $x'\in\Delta(x, u)$, which means that the system can evolve from the state $x$ to the state $x'$ under the input $u$. An input $u\in U$ belongs to \emph{the set of the enabled inputs} at the state $x$, denoted by $\enab(x)$, if $\Delta(x, u)\neq\varnothing$; see \cite{Girard2012controller}. If $\enab(x)=\varnothing$, then $x$ is said to be \emph{blocking} (or deadlock), otherwise, $x$ is said to be \emph{non-blocking}. The transition system $\mathbf{T}$ is said to be \emph{deterministic}, if for all $x\in X$ and all $u\in\enab(x)$, $\Delta(x, u)$ has exactly one element. In this case, let $x'=\Delta(x, u)$ with a slight abuse of notation.

\begin{definition}[\cite{Girard2007approximation}]
\label{def-2}
Let $\mathbf{T}_{i}=(X_{i}, X^{0}_{i}, U_{i}, \Delta_{i}, Y,$ $H_{i})$, $i=1, 2$, be two transition systems with the same output set $Y$ equipped with the metric $\mathbf{d}$. Given a precision  $\varepsilon>0$, a relation $\mathscr{R}\subseteq X_{1}\times X_{2}$ is called an \emph{$\varepsilon$-approximate bisimulation relation ($\varepsilon$-ABR)} between $\mathbf{T}_{1}$ and $\mathbf{T}_{2}$, denoted by $\mathbf{T}_{1}\simeq_{\varepsilon}\mathbf{T}_{2}$, if for all $(x_{1}, x_{2})\in\mathscr{R}$: (i) $\mathbf{d}(H_{1}(x_{1}), H_{2}(x_{2}))\leq\varepsilon$; (ii) for each $x'_{1}\in\Delta_{1}(x_{1}, u_{1})$ with $u_{1}\in\enab(x_{1})$, there exists $x'_{2}\in\Delta_{2}(x_{2}, u_{2})$ with $u_{2}\in\enab(x_{2})$ such that $(x'_{1}, x'_{2})\in\mathscr{R}$; (iii) for each $x'_{2}\in\Delta_{2}(x_{2}, u_{2})$ with $u_{2}\in\enab(x_{2})$, there exists $x'_{1}\in\Delta_{1}(x_{1}, u_{1})$ with $u_{1}\in\enab(x_{1})$ such that $(x'_{1}, x'_{2})\in\mathscr{R}$.
\end{definition}

\begin{definition}[\cite{Reissig2017feedback}]
\label{def-3}
Let $\mathbf{T}_{1}$ and $\mathbf{T}_{2}$ be two transition systems $\mathbf{T}_{i}=(X_{i}, X^{0}_{i}, U_{i}, \Delta_{i}, Y_{i}, H_{i})$ with $i\in\{1, 2\}$, and $U_{2}\subseteq U_{1}$. A relation $\mathscr{F}\subseteq X_{1}\times X_{2}$ is a \emph{feedback refinement relation (FRR)} from $\mathbf{T}_{1}$ to $\mathbf{T}_{2}$, denoted by $\mathbf{T}_{1}\preceq_{\mathscr{F}}\mathbf{T}_{2}$, if for all $(x_{1}, x_{2})\in\mathscr{F}$: $U_{2}(x_{2})\subseteq U_{1}(x_{1})$; $\mathscr{F}(\Delta_{1}(x_{1}, u))\subseteq\Delta_{2}(x_{2}, u)$ for all $u\in U_{2}(x_{2})$, where $U_{i}(x):=\{u\in U_{i}: u\in\enab(x)\}$.
\end{definition}

\begin{definition}[\cite{Girard2012controller}]
\label{def-4}
A \emph{controller} for a transition system $\mathbf{T}=(X, X_{0}, U, \Delta, Y, H)$ is a map $\mathscr{C}: X\rightarrow2^{U}$ with the domain defined as $\{x\in X: \mathscr{C}(x)\neq\varnothing\}$. It is \emph{well-defined}, if $\mathscr{C}(x)\subseteq\enab(x)$ for all $x\in X$. The \emph{controlled transition system} is denoted by the transition system $\mathbf{T}^{\textrm{c}}=(X, X_{0}, U, \Delta_{\textrm{c}}, Y, H)$, where $x'\in\Delta_{\textrm{c}}(x, u)$ if and only if $u\in\mathscr{C}(x)$ and $x'\in\Delta(x, u)$.
\end{definition}

\subsection{Linear Temporal Logic (LTL)}
\label{subsec-LTL}

Let $\mathcal{AP}$ be a set of atomic propositions \cite{Baier2008principles}. Based on atomic propositions (state labels $\alpha\in\mathcal{AP}$), Boolean connectors like negation $\neg$ and conjunction $\wedge$, and two temporal operators $\bigcirc$ (`next') and $\mathsf{U}$ (`until'), LTL is formed via the syntax below:
\begin{align}
\label{eqn-1}
\varphi::=\mathsf{true}\mid\alpha\mid\neg\varphi\mid\varphi_{1}\wedge\varphi_{2}\mid\bigcirc\varphi\mid\varphi_{1}\mathsf{U}\varphi_{2},
\end{align}
where $\varphi, \varphi_{1}, \varphi_{2}$ are LTL formulas. The disjunction $\vee$, and temporal operators $\diamondsuit$ (`eventually') and $\square$ (`always') can be derived as $\varphi_{1}\vee\varphi_{2}:=\neg(\neg\varphi_{1}\wedge\neg\varphi_{2})$, $\diamondsuit\varphi:=\mathsf{true}\mathsf{U}\varphi$ and $\square\varphi:=\neg\diamondsuit\neg\varphi$. Let $2^{\mathcal{AP}}$ be the power set of $\mathcal{AP}$. 

\begin{definition}[{\cite[Chapter 2.2.1]{Clarke2018handbook}}]
\label{def-5}
A \emph{Kripke structure} is of the form $(X, X_{0}, \Delta, \mathcal{L})$ with a state set $X$, an initial state set $X_{0}\subseteq X$, a transition relation $\Delta\subseteq X\times X$, and a labeling function $\mathcal{L}: X\rightarrow2^{\mathcal{AP}}$. 
\end{definition}

An infinite \emph{word} over $2^{\mathcal{AP}}$ is an infinite sequence $\sigma:=\sigma_{0}\sigma_{1}\ldots\in(2^{\mathcal{AP}})^{\omega}$, where $\omega$ denotes infinite repetition and $\sigma_{k}\in2^{\mathcal{AP}}$, $k\in\mathbb{N}$. For each $i\in\mathbb{N}$, let $\sigma|_{i}:=\sigma_{i}\sigma_{i+1}\ldots$. The semantics of LTL is defined as follows.
\begin{itemize}
  \item $\sigma\models\alpha$ if $\alpha\in\sigma_{1}$;
  \item $\sigma\models\neg\varphi$ if $\sigma\not\models\varphi$;
  \item $\sigma\models\varphi_{1}\wedge\varphi_{2}$ if $\sigma\models\varphi_{1}$ and $\sigma\models\varphi_{2}$;
  \item $\sigma\models\bigcirc\varphi$ if $\sigma|_{2}\models\varphi$;
  \item $\sigma\models\varphi_{1}\mathsf{U}\varphi_{2}$ if there exists $k\geq0$ such that $\sigma|_{i}\models\varphi_{1}$ for all $0\leq i<k$ and $\sigma|_k\models\varphi_{2}$.
\end{itemize}
The union of infinite words satisfying $\varphi$ is denoted as $\mathsf{Words}(\varphi)=\{\sigma\in(2^{\mathcal{AP}})^{\omega}: \sigma\models\varphi\}$. From \cite[Theorem 5.41]{Baier2008principles}, any LTL formula $\varphi$ can be translated into a nondeterministic B\"uchi automaton $\mathsf{B}$ over $2^{\mathcal{AP}}$, which is defined below. 

\begin{definition}
\label{def-6}
A \emph{nondeterministic B\"uchi automaton (NBA)} $\mathsf{B}$ over $2^{\mathcal{AP}}$ is a tuple $\mathsf{B}=(\mathcal{Q}_{\mathsf{B}}, \mathcal{Q}^{0}_{\mathsf{B}}, 2^{\mathcal{AP}}, \Delta_{\mathsf{B}}, \mathcal{F}_{\mathsf{B}})$, where $\mathcal{Q}_{\mathsf{B}}$ is a finite set of states, $\mathcal{Q}^{0}_{\mathsf{B}}\subseteq\mathcal{Q}_{\mathsf{B}}$ is the set of initial states, $2^{\mathcal{AP}}$ is the set of input alphabets, $\Delta_{\mathsf{B}}\subseteq\mathcal{Q}_{\mathsf{B}}\times2^{\mathcal{AP}}\times2^{\mathcal{Q}_{\mathsf{B}}}$ is a transition relation, and $\mathcal{F}_{\mathsf{B}}\subseteq\mathcal{Q}_{\mathsf{B}}$ is a set of accepting states.
\end{definition}

An infinite run $\mathsf{q}$ of $\mathsf{B}$ over an infinite word $\sigma=\sigma_{0}\sigma_{1}\ldots\in(2^{\mathcal{AP}})^{\omega}$ is a sequence $\mathsf{q}=\mathsf{q}_{0}\mathsf{q}_{1}\ldots$ such that $\mathsf{q}_{0}\in\mathcal{Q}^{0}_{\mathsf{B}}$ and $(\mathsf{q}_{k}, \sigma_{k}, \mathsf{q}_{k+1})\in\Delta_{\mathsf{B}}$, $k\in\mathbb{N}$. An infinite run $\mathsf{q}$ is called \emph{accepting}, if $\mathsf{Inf}(\mathsf{q})\cap\mathcal{F}_{\mathsf{B}}\neq\varnothing$, where $\mathsf{Inf}(\mathsf{q})$ is the set of states that appear in $\mathsf{q}$ infinitely often. The set of infinite words over $2^{\mathcal{AP}}$ resulting in an accepting run of $\mathsf{B}$ is called the accepted language of $\mathsf{B}$ and is denoted as $\mathfrak{L}(\mathsf{B})$. From \cite[Theorem 5.41]{Baier2008principles}, the NBA $\mathsf{B}$ associated with an LTL formula $\varphi$ does always exists such that $\mathsf{B}$ accepts all and only those infinite runs over $\mathcal{AP}$ that satisfy $\varphi$, i.e., $\mathfrak{L}(\mathsf{B})=\mathsf{Words}(\varphi)$.

\subsection{Nonlinear Control Systems}
\label{subsec-nonsystem}

Consider the nonlinear control system
\begin{equation}
\label{eqn-2}
\Sigma: \quad \dot{x}(t)=f(x(t), u(t)),
\end{equation}
where $x(t)\in\mathbb{X}$ is the system state, and $u(t)\in\mathbb{U}$ is the control input. $\mathbb{X}\subset\mathbb{R}^{n}$ and $\mathbb{U}\subseteq\mathbb{R}^{m}$ are assumed to be compact. The initial state set is denoted as $\mathbb{X}_{0}\subset\mathbb{X}$, and is assumed to be a constrained zonotope. The controller is assumed to be piecewise continuous, and the function $f: \mathbb{X}\times\mathbb{U}\rightarrow\mathbb{R}^{n}$ is assumed to be continuous and satisfy the local Lipschitz assumption for all $u\in\mathbb{U}$. For \eqref{eqn-2}, a curve $\mathbf{x}: \mathbb{R}^{+}\rightarrow\mathbb{X}$ is called a \emph{trajectory}, if there exists a control input $u(t)\in\mathbb{U}$ such that $\dot{\mathbf{x}}(t)=f(\mathbf{x}(t), u(t))$ for all $t\in\mathbb{R}^{+}$. We denote by $\mathbf{x}(t, x, u)$ the state reached at the time $t\in\mathbb{R}^{+}$ under the input $u\in\mathbb{U}$ from the initial state $x\in\mathbb{X}$. 

Using the sampling technique, we derive the time discretization of \eqref{eqn-2}. Let $\tau>0$ be the sampling period to be designed. The sampled-data system of \eqref{eqn-2} is denoted as a transition system $\mathbf{T}_{\tau}(\Sigma):=(X_{1}, X^{0}_{1}, U_{1}, \Delta_{1}, Y_{1}, H_{1})$, where, $X_{1}=\mathbb{X}$ is the state set; $X^{0}_{1}=\mathbb{X}_{0}$ is the set of initial states; $U_{1}=\{u\in\mathbb{U}: \mathbf{x}(\tau, x, u) \text{ is defined for all } x\in\mathbb{X}\}$ is the input set; $\Delta_{1}$ is the transition relation: for $x\in X_{1}$ and $u\in U_{1}$, $x'=\Delta_{1}(x, u)$ if and only if $x'=\mathbf{x}(\tau, x, u)$; $Y_{1}=X_{1}$ is the output set; and $H_{1}=\Id$ is the output map. The system $\mathbf{T}_{\tau}(\Sigma)$ is non-blocking and deterministic. $\mathbf{T}_{\tau}(\Sigma)$ is metric if the output set $Y_{1}$ is equipped with the metric $\textbf{d}(y, y')=\|y-y'\|$ for all $y, y'\in Y_{1}$. If the state space is limited to a subset $\mathbbmss{X}\subseteq\mathbb{X}$, then we use the notation $\mathbf{T}_{\tau}(\Sigma, \mathbbmss{X})$ to emphasize the local state space $\mathbbmss{X}\subseteq\mathbb{X}$.

\section{Problem Formulation}
\label{sec-problem}

In the paper, we consider the system \eqref{eqn-2} and aim to design a controller such that the desired specification described by an LTL formula $\varphi$ can be satisfied. For this purpose, we start with the high-level plan to satisfy the LTL formula $\varphi$. 

In the state space $\mathbb{X}\subset\mathbb{R}^{n}$, the set of all the obstacles is denoted by $\mathbb{O}:=\cup_{l\in\mathbb{K}}\mathbb{O}_{l}\subset\mathbb{X}$ with a finite index set $\mathbb{K}\subset\mathbb{N}^{+}$. We consider the existence of the set $\Pi\subseteq\mathcal{AP}$ of propositions for \eqref{eqn-2}; see \cite{Guo2015multi, Fainekos2009temporal, Meyer2019hierarchical}. Each $\pi_{i}\in\Pi$ is associated with a subset $\mathbb{X}_{i}\subseteq\mathbb{X}\setminus\mathbb{O}$ such that $\pi_{i}=\mathsf{true}$ if $x\in\mathbb{X}_{i}$, where $i\in\{0, 1, \ldots, |\Pi|\}$ and $|\Pi|$ is finite. In particular, $\mathbb{X}_{0}$ is the initial state set. In order to show the relation between $\pi_{i}\in\Pi$ and $\mathbb{X}_{i}\subseteq\mathbb{X}\setminus\mathbb{O}$, we denote $\mathbbmss{R}(\pi_{i}):=\mathbb{X}_{i}$. All these subsets are called the \emph{regions of interest}. Let each $\mathbb{X}_{i}$ be a constrained zonotope, which is a reasonable since any convex polytope can be represented as a constrained zonotope \cite{Scott2016constrained}.

Based on the above, we can define a Kripke structure $\mathsf{S}:=(\Pi, \Pi_{0}, \Delta_{\mathsf{S}}, \mathcal{L})$, where $\Pi$ is defined above, $\Pi_{0}\subseteq\Pi$, $\Delta_{\mathsf{S}}\subseteq\Pi\times\Pi$ is the transition relation and $\mathcal{L}: \Pi\rightarrow2^{\mathcal{AP}}$ is the labeling function. We define an infinite \emph{path} of $\mathsf{S}$ as an infinite sequence of states $\pi=\pi_{0}\pi_{1}\ldots$ such that $(\pi_{k}, \pi_{k+1})\in\Delta_{\mathsf{S}}$ for all $k\in\mathbb{N}$. A state sequence results in an infinite word  over $2^{\mathcal{AP}}$, which is defined as  $\mathsf{Word}(\pi):=\mathcal{L}(\pi_{0})\mathcal{L}(\pi_{1})\ldots$. Consider the LTL formula $\varphi$ and a path $\pi$, we have $\mathsf{Word}(\pi)\models\varphi$ if $\mathsf{Word}(\pi)\in\mathsf{Words}(\varphi)$. Given the Kripke structure $\mathsf{S}=(\Pi, \Pi_{0}, \Delta_{\mathsf{S}}, \mathcal{L})$ and the NBA $\mathsf{B}=(\mathcal{Q}_{\mathsf{B}}, \mathcal{Q}^{0}_{\mathsf{B}}, 2^{\mathcal{AP}}, \Delta_{\mathsf{B}}, \mathcal{F}_{\mathsf{B}})$, the product B\"uchi automaton is defined as $\mathsf{P}:=\mathsf{S}\times\mathsf{B}=(\mathcal{Q}_{\mathsf{P}}, \mathcal{Q}^{0}_{\mathsf{P}}, 2^{\mathcal{AP}}, \Delta_{\mathsf{P}}, \mathcal{F}_{\mathsf{P}})$, where $\mathcal{Q}_{\mathsf{P}}=\Pi\times\mathcal{Q}_{\mathsf{B}}$, $\mathcal{Q}^{0}_{\mathsf{P}}=\Pi_{0}\times\mathcal{Q}^{0}_{\mathsf{B}}$, $\mathcal{F}_{\mathsf{P}}=\Pi\times\mathcal{F}_{\mathsf{B}}$, and $\Delta_{\mathsf{P}}\subseteq\mathcal{Q}_{\mathsf{P}}\times\mathcal{Q}_{\mathsf{P}}$ defined as $((\pi_{\mathsf{a}}, \mathsf{q}_{\mathsf{a}}), (\pi_{\mathsf{b}}, \mathsf{q}_{\mathsf{b}}))\in\Delta_{\mathsf{P}}$ if and only if $\pi_{\mathsf{b}}=\Delta_{\mathsf{S}}(\pi_{\mathsf{a}})$ and $\mathsf{q}_{\mathsf{b}}=\Delta_{\mathsf{B}}(\mathsf{q}_{\mathsf{a}})$. Following \cite[Algorithm 1]{Lindemann2019coupled} and \cite[Algorithm 3]{Guo2015multi}, we can find a run of $\mathsf{P}$ defined as $\mathsf{p}:=\mathsf{p}_{0}\mathsf{p}_{1}\ldots=(\pi_{0}, \mathsf{q}_{0})(\pi_{1}, \mathsf{q}_{1})\ldots$ with $\pi=\pi_{0}\pi_{1}\ldots$ such that $\mathsf{Word}(\pi)\models\varphi$. In this case, the derived path $\pi=\pi_{0}\pi_{1}\ldots$ is called an \emph{accepting path}. From any accepting path $\pi$, the path-based region set is defined as
\begin{equation}
\label{eqn-3}
\mathbbmss{R}(\pi):=\{\mathbbmss{R}(\pi_{i})\subseteq\mathbb{X}\setminus\mathbb{O}: \pi_{i}\in\pi, i\in\mathbb{N}\}.
\end{equation}

\begin{figure}[!t]
\begin{center}
\begin{picture}(55, 80)
\put(-55, -14){\resizebox{55mm}{32mm}{\includegraphics[width=2.5in]{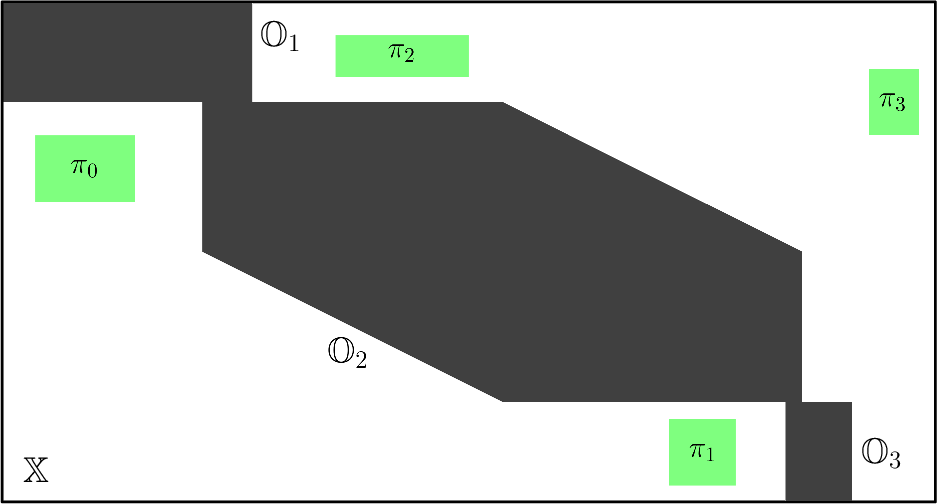}}}
\end{picture}
\end{center}
\caption{The state space of the robot in Example 1. The four regions of interest are denoted as $\mathbbmss{R}(\pi_{0}), \mathbbmss{R}(\pi_{1}), \mathbbmss{R}(\pi_{2}), \mathbbmss{R}(\pi_{3})$ with the propositions $\pi_{0}, \pi_{1}, \pi_{2}, \pi_{3}$, and the three obstacles are $\mathbb{O}_{1}, \mathbb{O}_{2}, \mathbb{O}_{3}$.}
\label{fig-1}
\end{figure}

\begin{example}
\label{expl-1}
Consider a robot which moves in a given state space $\mathbb{X}\subset\mathbb{R}^{2}$ and is initially placed in a certain region $\mathbb{X}_{0}\subset\mathbb{X}$. The robot is to accomplish the task: ``patrol between Regions A and B while visiting Region C not before Region B has been visited at least once'', which can be expressed as an LTL formula $\varphi=\square\diamondsuit\pi_{1}\wedge\square\diamondsuit\pi_{2}\wedge\diamondsuit\pi_{3}\wedge\neg\pi_{3}\mathsf{U}\pi_{2}$. Let $\mathcal{AP}=\{\pi_{0}, \pi_{1}, \pi_{2}, \pi_{3}\}$. The regions of interest are $\mathbbmss{R}(\pi_{0})=\mathbb{X}_{0}$ and $\mathbbmss{R}(\pi_{1}), \mathbbmss{R}(\pi_{2}), \mathbbmss{R}(\pi_{3})$, which are respectively Regions A, B and C; see Fig. \ref{fig-1}. We can derive an NBA $\mathsf{B}=(\mathcal{Q}_{\mathsf{B}}, \mathcal{Q}^{0}_{\mathsf{B}}, 2^{\mathcal{AP}}, \Delta_{\mathsf{B}}, \mathcal{F}_{\mathsf{B}})$ to be associated with $\varphi$. On the other hand, we can define a Kripke structure $\mathsf{S}=(\Pi, \Pi_{0}, \Delta_{\mathsf{S}}, \mathcal{L})$ by letting $\Pi=\mathcal{AP}$. We can derive the high-level plan via \cite[Algorithm 3]{Guo2015multi}, and obtain an accepting path $\pi=\pi_{0}\pi_{1}\pi_{2}\pi_{3}(\pi_{1}\pi_{2})^\omega$. This accepting path provides a guide for the controller synthesis afterwards. That is, from the accepting path $\pi$, we only need to deal with the following five pairs: $(\pi_{0}, \pi_{1}), (\pi_{1}, \pi_{2}), (\pi_{2}, \pi_{3}), (\pi_{3}, \pi_{1})$, and $(\pi_{2}, \pi_{1})$.
\hfill $\lhd$
\end{example}

Next, we need to confirm the realization of the derived accepting path and further design the controller such that the LTL specification is satisfied for the considered system. To this end, the following definition is introduced.

\begin{definition}
\label{def-7}
Consider the state space $\mathbb{X}\subset\mathbb{R}^{n}$, the obstacle set $\mathbb{O}\subseteq\mathbb{X}$, and the LTL formula $\varphi$ with its accepting path $\pi$. If there exists a connected set $\mathbbmss{X}\subseteq\mathbb{X}\setminus\mathbb{O}$ such that $\mathbbmss{X}\cap\mathbbmss{R}(\pi_{i})\neq\varnothing$ for all $\pi_{i}\in\pi$, then $\pi$ is said to be \emph{realized} in $\mathbb{X}$. The set $\mathbbmss{X}$ is called a \emph{realization region} of $\pi$.
\end{definition}

From Definition \ref{def-7}, the realization of $\pi$ shows that there exists a subset in $\mathbb{X}$ such that the system \eqref{eqn-2} is not thwarted to achieve the LTL specification $\varphi$. Note that the realization of the accepting path is necessary. If the accepting path is not realized, then it means that the LTL specification cannot be satisfied by any controller, which needs to be ruled out; see Fig. \ref{fig-1}. Since the realization of the accepting path is verified first, the need of the controller design can be checked and the cost for the controller design can be reduced. For instance, the accepting path in Example \ref{expl-1} is not realized in $\mathbb{X}$, since there does not exist a connect subset $\mathbbmss{X}\subseteq\mathbb{X}\setminus\mathbb{O}$ such that $\mathbbmss{X}\cap\mathbbmss{R}(\pi_{1})\neq\varnothing$ and $\mathbbmss{X}\cap\mathbbmss{R}(\pi_{2})\neq\varnothing$ hold simultaneously.

Based on the above discussion, the problems to be studied in this paper are formulated as follows.
\begin{itemize}\setlength{\itemindent}{-2em}
\item[] \textbf{Problem 1.} Given the system \eqref{eqn-2} and an LTL formula $\varphi$ with its accepting path $\pi$, verify whether $\pi$ is realized.
\item[] \textbf{Problem 2.} If $\pi$ is realized, then design a controller such that the LTL formula $\varphi$ is satisfied for the system  \eqref{eqn-2}.
\end{itemize}

\section{Zonotope-based Covering Approach}
\label{sec-partition}

To deal with \textbf{Problems 1} and \textbf{2}, we first implement zonotope techniques to propose a novel cover of the state space in this section. To be specific, we summarize the covering strategy in Section \ref{subsec-strategy}, then show the detailed generation of (constrained) zonotopes in Section \ref{subsec-generatezonotope}, and finally derive a graph from the proposed generation mechanism in Section \ref{subsec-topgraph}.

\subsection{Summary of Covering Strategy}
\label{subsec-strategy}

\begin{algorithm}[!t]
\DontPrintSemicolon\small
\caption{Cover of the State Space}
\label{alg-1}
\KwIn{$\mathbb{X}\subset\mathbb{R}^{n}, \varepsilon>0$}
\KwOut{the set $\mathbb{Z}$ of zonotopes and constrained zonotopes}
Generate finite zonotopes $\mathbf{Z}_{i}$ and constrained zonotopes $\mathbf{Z}^{\cstr}_{j}$ to cover $\mathbb{X}$\;
Expand both $\mathbf{Z}_{i}$ and $\mathbf{Z}^{\cstr}_{j}$ via the parameter $\epsilon>0$ as in \eqref{eqn-4} \;
\textbf{return} $\mathbb{Z}=(\cup^{N}_{i=1}\mathbf{E}_{\epsilon}(\mathbf{Z}_{i}))\cup(\cup^{M}_{j=1}\mathbf{E}_{\epsilon}(\mathbf{Z}^{\cstr}_{j}))$
\end{algorithm}

\begin{figure}[!t]
\begin{center}
\begin{picture}(75, 105)
\put(-75, -15){\resizebox{75mm}{45mm}{\includegraphics[width=2.5in]{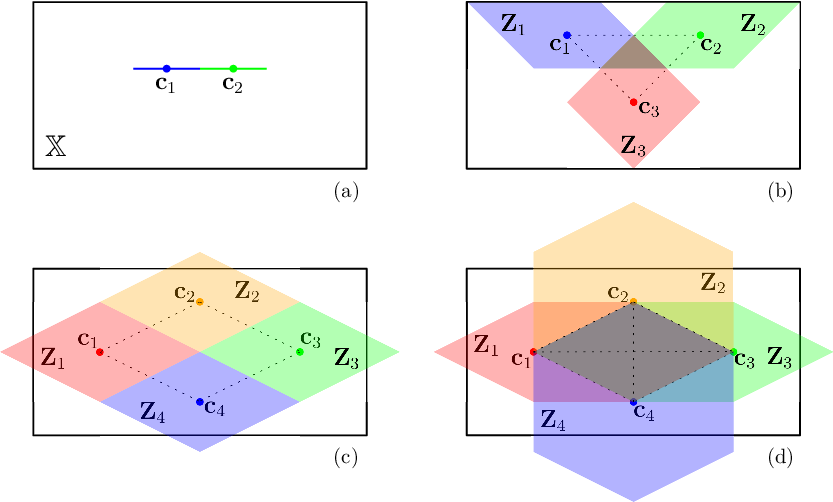}}}
\end{picture}
\end{center}
\caption{Illustration of the generation of zonotopes in different cases. (a) $N=2<3$: the generated zonotopes are two segments and thus not well-constructed. (b) $N=3$: three zonotopes are generated and overlapped. (c) $N=4$ and each center connects with 2 neighbor centers: four zonotopes are generated but not overlapped. (d) $N=4$ and each center connects with 3 neighbor centers: four zonotopes are generated and overlapped.}
\label{fig-2}
\end{figure}

The covering strategy is presented in Algorithm \ref{alg-1} and has two steps. The first step is to generate finite zonotopes and constrained zonotopes to cover the state space $\mathbb{X}\subset\mathbb{R}^{n}$ (i.e., line 1 in Algorithm \ref{alg-1}), and the detailed generation rules will be presented in the next subsection. Since all generated zonotopes and constrained zonotopes are not necessarily overlapped (see Fig. \ref{fig-2}(c)), the second step is to implement the expansion operator to expand all generated zonotopes and constrained zonotopes (i.e., line 2 in Algorithm \ref{alg-1}). Given any zonotope $\mathbf{Z}=\{\mathbf{c}, \mathbf{G}\}$ and any constrained zonotope $\mathbf{Z}^{\cstr}=\{\mathbf{c}, \mathbf{G}, \mathbf{A}, \mathbf{b}\}$,
\begin{align}
\label{eqn-4}
\begin{aligned}
\mathbf{E}_{\epsilon}(\mathbf{Z})&=\{\mathbf{c}, (1+\epsilon)\mathbf{G}\}, \\
\mathbf{E}_{\epsilon}(\mathbf{Z}^{\cstr})&=\{\mathbf{c}, (1+\epsilon)\mathbf{G}, \mathbf{A}, (1+\epsilon)\mathbf{b}\},
\end{aligned}
\end{align}
which are respectively treated as the $\epsilon$-expansions of $\mathbf{Z}$ and $\mathbf{Z}^{\cstr}$ with a slight abuse of notation. These expansion operations have explicit forms and ensures each (constrained) zonotope to overlap with its neighbor zonotopes and constrained zonotopes. Hence, Algorithm \ref{alg-1} produces the union of finite zonotopes and constrained zonotopes as in line 3. In the following, the generated zonotopes and constrained zonotopes are called the \emph{cells} if no confusion arises.

Different from existing partition approaches \cite{Belta2007symbolic, Fainekos2009temporal}, the proposed approach allows all cells to intersect with their neighbor cells, and we emphasize here that the intersection relation will play an important role in the controller synthesis afterwards. In particular, due to the intersection relation, the proposed approach results in an undirected graph, which will be applied to verify the desired specification in Section \ref{sec-verifydiscrete}, and the global specification can be decomposed into finite local ones such that local controllers can be designed individually to ensure the global specification in Section \ref{sec-abstractionbased}.

\begin{algorithm}[!t]
\DontPrintSemicolon\small
\caption{Zonotope Generation}
\label{alg-2}
\KwIn{$\mathbb{X}\subset\mathbb{R}^{n}$}
\KwOut{finite zonotopes and constrained zonotopes}
Choose $\{\mathbf{c}_{i}\in\mathbb{X}: i=1, \ldots, N\}$ with $N>n$\;
Connect these points such that for each $\mathbf{c}_{i}\in\mathbb{R}^{n}$, there exists a full-rank matrix
\begin{equation}
\label{eqn-5}
\mathbf{G}_{i}=(\mathbf{c}_{i_{1}}-\mathbf{c}_{i}, \ldots, \mathbf{c}_{i_{k}}-\mathbf{c}_{i}), \quad i_{k}\geq n
\end{equation}
where $i_{l}\in\{1, \ldots, N\}\setminus\{i\}$ and $l, k\in\mathbb{N}$\;
{With the center $\mathbf{c}_{i}\in\mathbb{R}^{n}$ and the generator matrix $0.5\mathbf{G}_{i}$, construct the zonotope
\begin{equation}
\label{eqn-6}
\mathbf{Z}_{i}=\{\mathbf{c}_{i}+0.5\mathbf{G}_{i}\xi: \|\xi\|\leq1\}
\end{equation}}\;
\vspace*{-\baselineskip}
\eIf{$\mathbb{X}\setminus(\cup^{N}_{i=1}\mathbf{Z}_{i})=\varnothing$}{
No need to construct constrained zonotopes
}{
Determine the intersection among $\partial\mathbf{Z}_{i}, \ldots, \partial\mathbf{Z}_{N}$ and $\partial\mathbb{X}$
\begin{equation}
\label{eqn-7}
\mathbb{H}_{1}=\{x\in\mathbb{X}: x\in\partial\mathbf{Z}_{i}\cap\partial\mathbf{Z}_{j} \text{ or } x\in\partial\mathbf{Z}_{i}\cap\partial\mathbb{X}\}
\end{equation}
where $i, j\in\{1, \ldots, N\}$ and $i\neq j$\;
Refine the set $\mathbb{H}_{1}$ into the following set
\begin{equation}
\label{eqn-8}
\mathbb{H}=\{x\in\mathbb{H}_{1}: x\notin\mathbf{Z}^{\circ}_{i}\}
\end{equation}\;
\vspace*{-\baselineskip}
Connect all elements in $\mathbb{H}$ to generate a finite set $\mathbb{S}$ of convex regions such that
\begin{equation}
\label{eqn-9}
\mathbb{X}\setminus(\cup^{N}_{i=1}\mathbf{Z}_{i})\subseteq\mathbb{S} \text{ and } \mathbb{S}\cap(\cup^{N}_{i=1}\mathbf{Z}^{\circ}_{i})=\varnothing
\end{equation}\;
\vspace*{-\baselineskip}
Construct $M=|\mathbb{S}|$ constrained zonotopes $\{\mathbf{Z}^{\cstr}_{j}: j=1, \ldots, M\}$ such that
\begin{equation}
\label{eqn-10}
\cup^{M}_{j=1}\mathbf{Z}^{\cstr}_{j}\supseteq\mathbb{S}
\end{equation}}
\textbf{return} $(\cup^{N}_{i=1}\mathbf{Z}_{i})\cup(\cup^{M}_{j=1}\mathbf{Z}^{\cstr}_{j})$
\end{algorithm}

\subsection{Generation of Zonotopes and Constrained Zonotopes}
\label{subsec-generatezonotope}

In this subsection, we show how to generate zonotopes and constrained zonotopes, presented in Algorithm \ref{alg-2}. The generation rule consists of two parts: the first part is to generate zonotopes (lines 1-5), and the second part is to generate constrained zonotopes (lines 6-10).

To begin with, the number of zonotopes to be generated is set \emph{a priori} as $N\in\mathbb{N}$, and we choose $N$ points $\mathbf{c}_{i}\in\mathbb{X}$ as the centers of zonotopes, where $i\in\{1, \ldots, N\}$. Here, we assume $N>n$, where $n\in\mathbb{N}$ is the state-space dimension. Since only the number $N$ is constrained, there is no constraint on how to choose these points, which thus can be chosen arbitrarily. Second, we connect these centers such that each center is connected with at least $n\in\mathbb{N}$ neighbor centers. Therefore, for each center, these connections leads to at least $n\in\mathbb{N}$ vectors, which will be further used as the generators for each zonotope. Hence, the constraints on the number $N\in\mathbb{N}$ and the matrix $\mathbf{G}_{i}$ are to ensure the construction of zonotopes, which is illustrated in Fig. \ref{fig-2}. Finally, with these centers and generators, we generate a zonotope $\mathbf{Z}_{i}$ as in line 3 of Algorithm \ref{alg-2}. In \eqref{eqn-6}, the coefficient $0.5$ is to guarantee each generated zonotope to be as large as possible and the intersection region of neighbor zonotopes to be as small as possible.

The generation rule of zonotopes is illustrated in Fig. \ref{fig-2}. Given a 2-dimensional space $\mathbb{X}\subset\mathbb{R}^{2}$, if $N=2$ and we choose two centers $\mathbf{c}_{1}, \mathbf{c}_{2}$ as in Fig. \ref{fig-2}(a), then the generated zonotopes are two one-dimensional segments, which cannot be used in the analysis afterwards, implying that the zonotopes are not well-constructed. If the matrix $\mathbf{G}_{i}$ is not full-rank, then similar cases may occur, where the zonotopes are not well-constructed. Therefore, the constraints on $N\in\mathbb{N}$ and the matrix $\mathbf{G}_{i}$ are necessary for the well-constructedness of all generated zonotopes. In addition, different choices of $N\in\mathbb{N}$ and $i_{k}\in\mathbb{N}$ in \eqref{eqn-5} have effects on the generated zonotopes and result in different zonotopes; see Figs. \ref{fig-2}(b)-(d).

\begin{figure}[!t]
\begin{center}
\begin{picture}(70, 130)
\put(-70, -15){\resizebox{70mm}{50mm}{\includegraphics[width=2.5in]{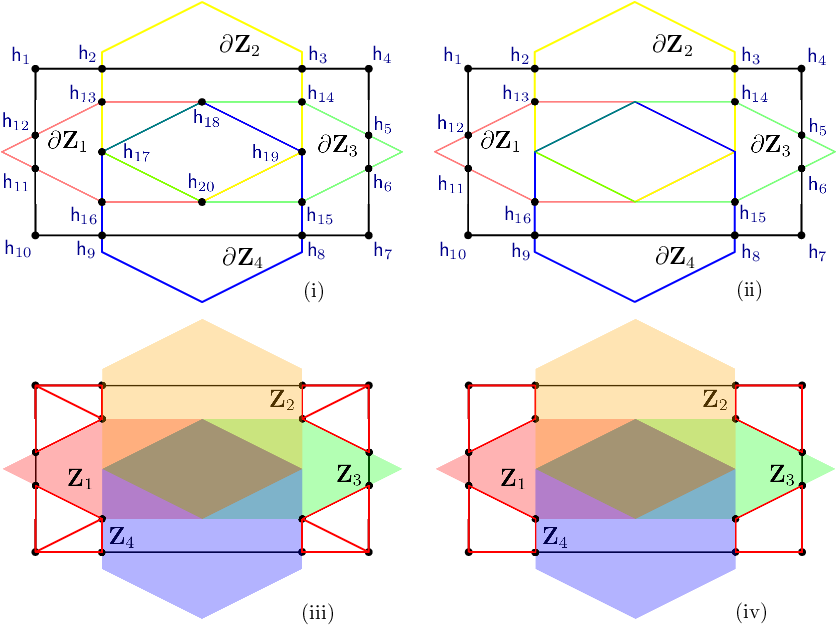}}}
\end{picture}
\end{center}
\caption{Illustration of the generation of constrained zonotopes for the case in Fig. \ref{fig-2}(d). (i) Illustration of line 7 in Algorithm \ref{alg-2}. The boundaries of four zonotopes are lines in different colors, and $\mathbb{H}_{1}=\{\mathsf{h}_{1}, \ldots, \mathsf{h}_{20}\}$. (ii) Illustration of line 8, and $\mathbb{H}=\{\mathsf{h}_{1}, \ldots, \mathsf{h}_{16}\}$. (iii)-(iv) Illustration of different ways to lines 9-10. In (iii), the set $\mathbb{S}$ has 8 triangle regions, which are used to construct $M=8$ constrained zonotopes. In (iv), the set $\mathbb{S}$ has 4 regions, which are used to construct $M=4$ constrained zonotopes.}
\label{fig-3}
\end{figure}

If the union of all generated zonotopes covers the state space, then there is no need for the generation of constrained zonotopes (lines 4-5); otherwise, the generation of constrained zonotopes is needed as in lines 7-10. The motivation of using constrained zonotopes lies in that constrained zonotopes are asymmetric \cite{Scott2016constrained, Rego2020guaranteed} and can thus be used to cover asymmetric regions. The generation rule of constrained zonotopes is presented below in detail. In line 7, we determine the intersection among boundaries of all generated zonotopes and the state space $\mathbb{X}$ to derive the set $\mathbb{H}_{1}$. In line 8, the set $\mathbb{H}_{1}$ is refined into the set $\mathbb{H}$ by ruling out the elements which are in both $\mathbb{H}_{1}$ and $\mathbf{Z}^{\circ}_{i}$ with certain $i\in\{1, \ldots, N\}$. In line 9, all elements are connected to generate finite convex regions to partition $\mathbb{X}\setminus(\cup^{N}_{i=1}\mathbf{Z}_{i})$. In line 10, these convex regions are represented as constrained zonotopes. See Fig. \ref{fig-3} for the illustration of the generation of constrained zonotopes.

The generation of constrained zonotopes is based on basic operations of polytopes. In line 7, the vertices of all zonotopes are obtained via the transformation of zonotopes from the G-representation into the V-representation \cite[Algorithm 2]{Kochdumper2019representation}. The V-representation determines the boundaries of all zonotopes and the set $\mathbb{H}_{1}$ is derived by verifying the intersection relation of all boundaries. In line 8, the refinement of $\mathbb{H}_{1}$ is based on Lemma \ref{lem-1}. Based on the vertices in $\mathbb{H}$ and using the transformation from V-representation into Z-representation \cite[Algorithm 1]{Kochdumper2019representation}, constrained zonotopes are generated in lines 9-10 to partition the region $\mathbb{X}\setminus(\cup^{N}_{i=1}\mathbf{Z}_{i})$. Note that other transformation approaches (see e.g., \cite{Scott2016constrained}) can be applied to show the reasonability of the generation mechanism.

The overall generation rule is summarized in Algorithm \ref{alg-2}, which produces finite zonotopes and constrained zonotopes (line 11) to cover the state space. After the expansion operation \eqref{eqn-4}, Algorithm \ref{alg-1} produces the set $\mathbb{Z}$. To simplify the following notations, all cells in $\mathbb{Z}$ are labeled via a finite symbol set $\mathbb{V}:=\{\mathsf{v}_{1}, \ldots, \mathsf{v}_{N+M}\}$, where $\mathbbmss{R}(\mathsf{v}_{i})=\mathbf{E}_{\epsilon}(\mathbf{Z}_{i})$ with $i\in\{1, \ldots, N\}$ and $\mathbbmss{R}(\mathsf{v}_{N+j})=\mathbf{E}_{\epsilon}(\mathbf{Z}^{\cstr}_{j})$ with $j\in\{1, \ldots, M\}$. Let $\mathcal{N}:=\{1, \ldots, N+M\}$, and the cover of $\mathbb{X}$ is $\mathscr{P}(\mathbb{X})=\mathbb{Z}$ with
\begin{align}
\label{eqn-11}
\mathbb{Z}:=\{\mathbb{Z}_{k}=\mathbbmss{R}(\mathsf{v}_{k}): k\in\mathcal{N}\}.
\end{align}

\begin{remark}
\label{rmk-1}
In Algorithm \ref{alg-2}, the number $N$ and the matrix $\mathbf{G}_{i}$ are not fixed, which indicates the flexibility of the proposed covering strategy. Different choices of $N$ and $\mathbf{G}_{i}$ have no effects on the feasibility of Algorithm \ref{alg-2}. That is, for fixed $N$ and $\mathbf{G}_{i}$, Algorithm \ref{alg-2} produces finite zonotopes and constrained zonotopes to cover the state space. However, the choices of $N$ and $\mathbf{G}_{i}$ affect the generation of zonotopes (see Fig. \ref{fig-2}), and further affect the computational complexity of Algorithm \ref{alg-2}. In particular, lines 7-9 involve basic operations of polytopes, which can be achieved efficiently with tools from  \cite{Kochdumper2019representation}. If $N$ and $\mathbf{G}_{i}$ are such that the constrained zonotopes are not needed (i.e., line 5 in Algorithm \ref{alg-2}), then the computational complexity of Algorithm \ref{alg-2} can be reduced.
\hfill $\vartriangle$
\end{remark}

\begin{algorithm}[!t]
\DontPrintSemicolon\small
\caption{Adjacency Matrix}
\label{alg-3}
\KwIn{$\mathbb{Z}$, $\mathbb{O}\subset\mathbb{X}$, $\varepsilon\in(0, \epsilon]$}
\KwOut{$A=[a_{ij}], \mathbbmss{I}=\{\mathbbmss{I}_{ij}\}$}
\For{$i=1:1:N+M-1$}{
\eIf{$\mathbb{Z}^{\varepsilon}_{i}\cap\mathbf{O}\neq\mathbb{Z}^{\varepsilon}_{i}$}{
\For{$j=1:1:N+M$}{
\eIf{$\mathbb{Z}^{\varepsilon}_{j}\cap\mathbf{O}\neq\mathbb{Z}^{\varepsilon}_{j}$}{
$\Omega=\mathbb{Z}^{\varepsilon}_{i}\cap\mathbb{Z}^{\varepsilon}_{j}$\;
\eIf{$\Omega=\varnothing$}{
$a_{ij}=0$\;}{
\eIf{$(\mathbb{Z}^{\varepsilon}_{i}\cup\mathbb{Z}^{\varepsilon}_{j})\setminus(\Omega\cap\mathbf{O})$ is connected}{
$a_{ij}=1$\;
$\mathbbmss{I}_{ij}=\Omega\setminus\mathbf{O}$\;
}{$a_{ij}=0$\;}
}
}{$a_{ij}=0$\;}
}
}{$a_{ij}=0$ for $j\in\mathcal{N}$\;}
}
\textbf{return} $A=[a_{ij}], \mathbbmss{I}=\{\mathbbmss{I}_{ij}\}$
\end{algorithm}

\begin{remark}
\label{rmk-2}
As mentioned in Section \ref{subsec-strategy}, each cell intersects with its neighbor cells. From Figs. \ref{fig-1}(b)-\ref{fig-1}(d), the intersection comes partially and naturally from the generation rule, which is further enhanced via line 2 in Algorithm \ref{alg-1}. Although other classes of polytopes \cite{Belta2007symbolic, Gol2013time} have been applied to cover the state space, the properties of zonotopes and constrained zonotopes will be involved in the controller design afterwards, which is not the case in \cite{Belta2007symbolic, Gol2013time}.
\hfill $\vartriangle$
\end{remark}

\subsection{Topological Graph}
\label{subsec-topgraph}

In this subsection, we follow the intersection relation among all cells to derive an undirected graph. For this purpose, some auxiliary notations are introduced. Given two regions $\mathbb{A}, \mathbb{B}\subset\mathbb{X}$, their intersection $\mathbb{A}\cap\mathbb{B}$ is \emph{admissible}, if $\mathbb{A}\cup\mathbb{B}\setminus(\mathbb{A}\cap\mathbb{B}\cap\mathbb{O})$ is connected. Given an $\varepsilon\in(0, \epsilon]$, we denote $\mathbf{O}:=\mathbf{E}_{\varepsilon}(\mathbb{O})$.

To derive the graph, the key is how to determine the existence of an edge between two neighbor cells, which can be achieved by verifying whether all intersection regions are admissible. For instance, if the intersection region between two neighbor cells belongs to the obstacle set, then this intersection region is not admissible and no edge is defined between these two neighbor cells. By verifying all intersection regions, we have all admissible intersection regions and the adjacency matrix, which is summarized in Algorithm \ref{alg-3}. In Algorithm \ref{alg-3}, the contraction and expansion operations in lines 2-17 are implemented to facilitate the controller design afterwards. These contraction and expansion operations can be removed if we only focus on the realization of the accepting path.

\begin{lemma}
\label{lem-2}
Given the cover $\mathbb{Z}$ in \eqref{eqn-11} of the state space $\mathbb{X}\subseteq\mathbb{R}^{n}$, the adjacency matrix $A$ from Algorithm \ref{alg-3} shows all edges among all cell pairs.
\end{lemma}

\begin{figure}[!t]
\begin{center}
\begin{picture}(77, 80)
\put(-75, -12){\resizebox{77mm}{31mm}{\includegraphics[width=2.5in]{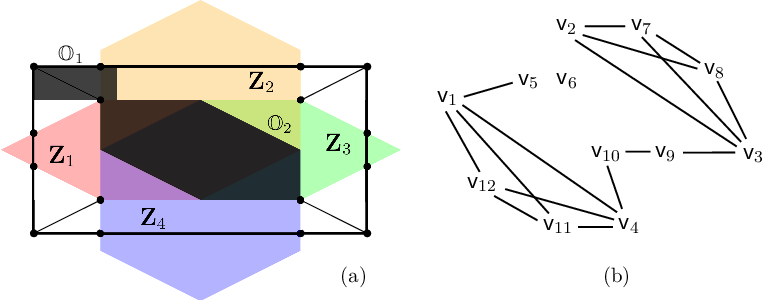}}}
\end{picture}
\end{center}
\caption{Illustration of Algorithm \ref{alg-3} via Fig. \ref{fig-3}(iii). (a) The cover of the state space $\mathbb{X}\subset\mathbb{R}^{2}$, and $\mathbb{O}_{3}$ in Fig. \ref{fig-1} is removed here. (b) The graph $\mathcal{G}=(\mathcal{V}, \mathcal{E})$ with $\mathcal{V}=\{\mathsf{v}_{1}, \ldots, \mathsf{v}_{12}\}$ and $\mathcal{E}$ from the adjacency matrix $A$.}
\label{fig-4}
\end{figure}

Lemma \ref{lem-2} is derived directly from Algorithm \ref{alg-3}. Due to the finite number of all cells, the termination time of Algorithm \ref{alg-3} is finite. From the graph theory, the matrix $A$ is further transformed into an undirected graph $\mathcal{G}=(\mathcal{V}, \mathcal{E})$, where the vertex set is $\mathcal{V}=\mathbb{V}$ and the edge set is $\mathcal{E}\subseteq\mathbb{V}\times\mathbb{V}$ with $(\mathsf{v}_{i}, \mathsf{v}_{j})\in\mathcal{E}$ if $a_{ij}=1$. See Fig. \ref{fig-4} for the illustration of Algorithm \ref{alg-3} and the generated graph.

\section{Realization Verification}
\label{sec-verifydiscrete}

Based on the cover of the state space, \textbf{Problem 1} is solved via the verification of the accepting path of the LTL specification in this section. To be specific, the graph-based approach is presented in Section \ref{subsec-graphbased} to verify the realization of the accepting path. The realization verification results in finite local LTL formulas in Section \ref{subsec-LocalLTL}.

\subsection{Graph-based Verification}
\label{subsec-graphbased}

To realize the accepting path $\pi$ of the LTL formula $\varphi$, it suffices to show that any pair $(\pi_{i}, \pi_{i+1})$ from $\pi$ can be connected. For this purpose and to facilitate the controller synthesis afterwards, we consider an auxiliary path $\pi_{\varepsilon}$, which is derived from the accepting path $\pi$ and $\varepsilon>0$ given in Algorithm \ref{alg-3}. Given the accepting path $\pi=\pi_{0}\pi_{1}\pi_{2}\ldots$, the auxiliary path is defined as $\pi_{\varepsilon}:=\pi^{\varepsilon}_{0}\pi^{\varepsilon}_{1}\pi^{\varepsilon}_{2}\ldots$ such that $\mathbbmss{R}(\pi^{\varepsilon}_{i})=\mathbbmss{R}^{\varepsilon}(\pi_{i})$ and $i\in\mathbb{N}$. That is, for each $i\in\mathbb{N}$, $\mathbbmss{R}(\pi^{\varepsilon}_{i})$ is the $\varepsilon$-contraction of $\mathbbmss{R}^{\varepsilon}(\pi_{i})$. Due to the contraction operation, the auxiliary path $\pi_{\varepsilon}$ is called a \emph{robust accepting path} of $\varphi$. 

\begin{figure}[!t]
\begin{center}
\begin{picture}(55, 70)
\put(-55, -14){\resizebox{55mm}{32mm}{\includegraphics[width=2.5in]{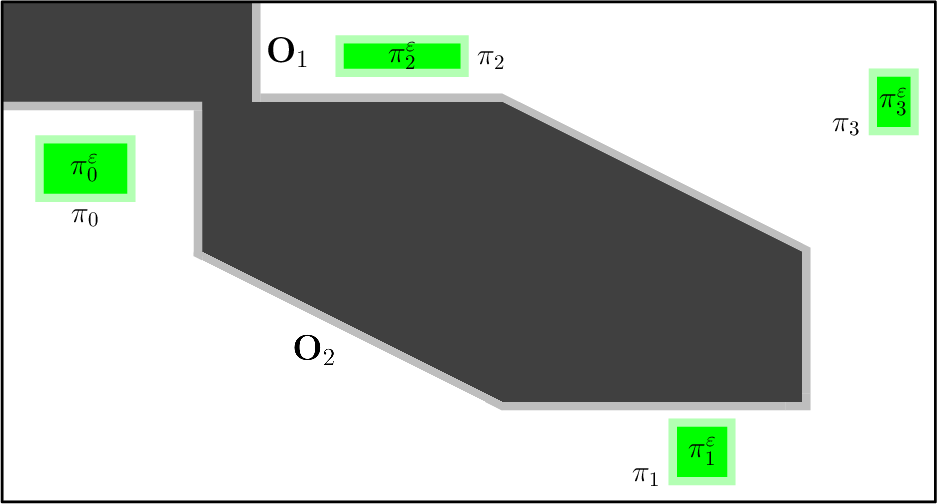}}}
\end{picture}
\end{center}
\caption{Illustration of the state space in Fig. \ref{fig-4}(a). For each $i\in\{0, 1, 2, 3\}$, the light green region is included in $\mathbbmss{R}(\pi_{i})$ while not in $\mathbbmss{R}(\pi^{\varepsilon}_{i})$. For each $j\in\{1, 2\}$, $\mathbf{O}_{j}=\mathbf{E}_{\varepsilon}(\mathbb{O}_{j})$ includes both dark and light grey regions.}
\label{fig-5}
\end{figure}

\begin{example}
\label{expl-2}
Let us revisit Example \ref{expl-1}. For the LTL formula $\varphi=\square\diamondsuit\pi_{1}\wedge\square\diamondsuit\pi_{2}\wedge\diamondsuit\pi_{3}\wedge\neg\pi_{3}\mathsf{U}\pi_{2}$, the accepting path is $\pi=\pi_{0}\pi_{1}\pi_{2}\pi_{3}(\pi_{1}\pi_{2})^\omega$. The robust accepting path is defined as $\pi_{\varepsilon}=\pi^{\varepsilon}_{0}\pi^{\varepsilon}_{1}\pi^{\varepsilon}_{2}\pi^{\varepsilon}_{3}(\pi^{\varepsilon}_{1}\pi^{\varepsilon}_{2})^\omega$. For each $i\in\{0, 1, 2, 3\}$, $\mathbbmss{R}(\pi^{\varepsilon}_{i})=\mathbbmss{R}^{\varepsilon}(\pi_{i})$, which is shown in Fig. \ref{fig-5}. 
\hfill $\lhd$
\end{example}

\begin{lemma}
\label{lem-3}
Given the LTL formula $\varphi$ with the accepting path $\pi$ and $\varepsilon>0$, if the robust accepting path $\pi_{\varepsilon}$ is realized, then the accepting path $\pi$ is realized.
\end{lemma}

\begin{IEEEproof}
From Definition \ref{def-7}, the realization of the accepting path $\pi_{\varepsilon}$ means that there exists a connected set $\mathbbmss{X}\subseteq\mathbb{X}\setminus\mathbb{O}$ such that $\mathbbmss{X}\cap\mathbbmss{R}(\pi^{\varepsilon}_{i})\neq\varnothing$ for all $\pi^{\varepsilon}_{i}\in\pi_{\varepsilon}$. From the relation between $\pi$ and $\pi_{\varepsilon}$, $\mathbbmss{R}(\pi_{\varepsilon})$ is an $\varepsilon$-contraction of $\mathbbmss{R}(\pi)$, and thus $\mathbbmss{X}\cap\mathbbmss{R}(\pi_{i})\neq\varnothing$. That is, for all $\pi_{i}\in\pi$, it holds $\mathbbmss{X}\cap\mathbbmss{R}(\pi_{i})\neq\varnothing$, which implies the realization of the accepting path $\pi$.
\end{IEEEproof}

Lemma \ref{lem-3} shows the relation between the realization of $\pi$ and $\pi_{\varepsilon}$, and next we focus on the robust accepting path $\pi_{\varepsilon}$. To verify the realization of $\pi_{\varepsilon}$, we generalize the graph $\mathcal{G}$ by including all symbols $\pi^{\varepsilon}_{i}$, $i\in\{0, 1, \ldots, |\pi_{\varepsilon}|\}$. To show this, the vertex set is generalized as $\overbar{\mathcal{V}}=\mathbb{V}\cup\{\pi^{\varepsilon}_{i}: i\in\{0, 1, \ldots, |\pi_{\varepsilon}|\}\}$. Since $\mathbbmss{R}(\pi_{\varepsilon})$ may intersect with some cells in $\mathbb{Z}$, these intersection relations can be checked and included into the generalized edge set $\bar{\mathcal{E}}\subseteq\overbar{\mathcal{V}}\times\overbar{\mathcal{V}}$. Hence, the generalized graph is $\bar{\mathcal{G}}=(\overbar{\mathcal{V}}, \bar{\mathcal{E}})$,  and each path $\pmb{\mathbbmss{p}}$ in $\bar{\mathcal{G}}$ results in a sequence of finite cells denoted as $\mathbbmss{P}:=\{\mathbbmss{P}_{k}=\mathbbmss{R}(\mathsf{v}_{k})\in\mathbb{Z}: \mathsf{v}_{k}\in\pmb{\mathbbmss{p}}, k\in\mathbb{N}^{+}\}$. The intersection region set in $\mathbbmss{P}$ is denoted as $\mathbbmss{I}:=\{\mathbbmss{I}_{ij}=\mathbbmss{P}_{i}\cap\mathbbmss{P}_{j}: \forall \mathbbmss{P}_{i}, \mathbbmss{P}_{j}\in\mathbbmss{P}\}$. From the graph $\bar{\mathcal{G}}$, the next theorem is derived to verify the realization of $\pi_{\varepsilon}$.

\begin{theorem}
\label{thm-1}
Consider the state space $\mathbb{X}\subseteq\mathbb{R}^{n}$, the LTL formula $\varphi$ with the accepting path $\pi_{\varepsilon}$ and a given precision $\varepsilon>0$. Let the generalized graph $\bar{\mathcal{G}}$ be defined as above. the robust accepting path $\pi_{\varepsilon}$ is realized in $\mathbb{X}$, if and only if
\begin{enumerate}[(i)]
  \item there exists a path $\pmb{\mathbbmss{p}}$ in $\bar{\mathcal{G}}$ such that $\pi_{\varepsilon}$ is embedded, that is, there exits at least one sub-path in $\pmb{\mathbbmss{p}}$ such that for any $k\in\{0, 1, \ldots, |\pi_{\varepsilon}|-1\}$,  $\pi^{\varepsilon}_{k}$ and $\pi^{\varepsilon}_{k+1}$ are connected;

  \item for each $\mathbbmss{P}_{i}\in\mathbbmss{P}$ with $\mathbbmss{P}$ from the path $\pmb{\mathbbmss{p}}$,
  \begin{enumerate}[({ii}-a)]
    \item $\mathbbmss{P}^{\varepsilon}_{i}\setminus\mathbf{O}$ is a connected region;

    \item otherwise, there exists a subregion $\mathbbmss{P}^{\prime}_{i}\subset\mathbbmss{P}^{\varepsilon}_{i}\setminus\mathbf{O}$ such that $\mathbbmss{P}^{\prime}_{i}\cup\mathbbmss{I}^{\varepsilon}_{(i-1)i}\cup\mathbbmss{I}^{\varepsilon}_{i(i+1)}$ is connected, and $\mathbbmss{P}^{\prime}_{i}\cap\mathbbmss{R}(\pi_{\varepsilon})\neq\varnothing$ if $\mathbbmss{P}_{i}\cap\mathbbmss{R}(\pi_{\varepsilon})\neq\varnothing$.
  \end{enumerate}
\end{enumerate}
\end{theorem}

\begin{IEEEproof}
See Appendix \ref{prf-thm1}.
\end{IEEEproof}

In Theorem \ref{thm-1}, item (i) implies that any pair $(\pi^{\varepsilon}_{i}, \pi^{\varepsilon}_{i+1})$ in $\pi_{\varepsilon}$ is connected in a sub-path $\pmb{\mathbbmss{p}}_{i}\subseteq\pmb{\mathbbmss{p}}$. Item (ii) is to verify the connectedness property of each cell in $\mathbbmss{P}$ from $\pmb{\mathbbmss{p}}$. For each cell in $\mathbbmss{P}$, it either is itself connected or has a connected subregion to intersect the previous and next cells. From Theorem \ref{thm-1}, the sequence of finite cells from the path $\pmb{\mathbbmss{p}}$ is denoted as
\begin{align}
\label{eqn-12}
\mathbbmss{P}:=\{\mathbbmss{P}_{i}=\mathbbmss{R}(\mathsf{v}_{i})\in\mathbb{Z}: \mathsf{v}_{i}\in\pmb{\mathbbmss{p}}\subset\overbar{\mathcal{V}}\}.
\end{align}
Similarly, $\mathbbmss{P}(\pi^{\varepsilon}_{i}, \pi^{\varepsilon}_{i+1})$ is the sequence of finite cells from the sub-path $\pmb{\mathbbmss{p}}_{i}$ for a pair $(\pi^{\varepsilon}_{i}, \pi^{\varepsilon}_{i+1})$. In particular, if there exist loops in $\pi_{\varepsilon}$, then there exists loops in $\pmb{\mathbbmss{p}}$ and each loop can be divided into two sub-sequences in $\mathbbmss{P}$. If a loop in $\pi_{\varepsilon}$ appears finite/infinite times, then the corresponding two sub-sequences in $\mathbbmss{P}$ only need to be checked once via item (ii).

Theorem \ref{thm-1} offers a graph-based approach to verify the realization of the accepting path and thus solves \textbf{Problem 1}. Note that the path $\pmb{\mathbbmss{p}}$ is not necessarily unique. Similar to Algorithm \ref{alg-3}, the contraction operation is applied to facilitate the controller design in Section \ref{sec-abstractionbased}, and can be reduced if only the realization is addressed.

\begin{remark}
\label{rmk-3}
The computational complexity of the realization verification essentially depends on the number of all cells. In terms of graph searching, the time for searching an admissible path depends on the number of all cells. In terms of graph generation, the existence of all edges lies in Algorithm \ref{alg-3} to verify the admissibility of all intersection regions, which has the computational complexity $\mathcal{O}((N+M)n^{3})$; see \cite{Scott2016constrained, Kochdumper2019representation}. Therefore, with an increasing number of cells, more operations are needed and the computational complexity increases. We stress that the realization verification is a premise for the control design, and thus its computational complexity is not included in the following sections.
\hfill $\vartriangle$
\end{remark}

\subsection{Decomposition into Local LTL Formulas}
\label{subsec-LocalLTL}

Once the accepting path of the LTL formula can be realized, the obtained path $\pmb{\mathbbmss{p}}$ can be projected into a sequence of cells in \eqref{eqn-12}. In this subsection, we formulate a local LTL formula for each cell in the derived sequence. 

From Section \ref{sec-partition}, any two neighbor cells overlap such that it is possible for the considered system to move between these two cells. Given the LTL formula $\varphi$, let $\mathbbmss{P}$ be the sequence of the cells derived from Theorem \ref{thm-1}, and $\mathsf{L}:=|\mathbbmss{P}|$. For each cell $\mathbbmss{P}^{\varepsilon}_{i}\in\mathbbmss{P}^{\varepsilon}$ with $i\in\mathbbmss{L}=\{0, \ldots, \mathsf{L}\}$, the following three types of local regions are derived.

\begin{enumerate}[(1)]
\item  For the first cell $\mathbbmss{P}_{0}$, $\mathbbmss{P}^{\varepsilon}_{0}\setminus\mathbf{O}$ is the local initial region in $\mathbbmss{P}_{0}$. For $i\in\mathbb{N}^{+}$, the local initial region is $(\mathbbmss{P}^{\varepsilon}_{i-1}\cap\mathbbmss{P}^{\varepsilon}_{i})\setminus\mathbf{O}$.

\item  For each $i\in\mathbb{N}$, if $\mathbbmss{P}^{\varepsilon}_{i+1}$ does not exist, then there exists no local target region in $\mathbbmss{P}^{\varepsilon}_{i}$. If $\mathbbmss{P}^{\varepsilon}_{i+1}$ exists, then $\mathscr{T}_{i}:=(\mathbbmss{P}^{\varepsilon}_{i}\cap\mathbbmss{P}^{\varepsilon}_{i+1})\setminus\mathbf{O}$ is the local target region in $\mathbbmss{P}^{\varepsilon}_{i}$.

\item  If $\mathbbmss{P}^{\varepsilon}_{i}\cap\mathbbmss{R}(\pi_{\varepsilon})\neq\varnothing$ and $\mathbbmss{P}^{\varepsilon}_{i}\cap\mathbbmss{P}^{\varepsilon}_{i+1}\cap\mathbbmss{R}(\pi_{\varepsilon})=\varnothing$, then all regions in $\mathbbmss{P}^{\varepsilon}_{i}\cap\mathbbmss{R}(\pi_{\varepsilon})$ are called \emph{local internal regions}, whose number is denoted as $\ell_{i}\in\mathbb{N}$ and which are denoted as $\mathscr{I}_{il}$ with $l\in\mathbb{L}_{i}:=\{1, \ldots, \ell_{i}\}$.
\end{enumerate}
Observe that the local target region of the current cell will be the local initial region of the next cell (if it exists). By verifying whether $\mathbbmss{P}^{\varepsilon}_{i}\cap\mathbbmss{R}(\pi_{\varepsilon})$ is empty, we can establish all local internal regions in $\mathbbmss{P}^{\varepsilon}_{i}$. The local target and internal regions depend on the LTL formula $\varphi$ and thus do not necessarily exist. If they are existent, then the local target region may intersect with local internal regions, that is, $\mathbbmss{P}^{\varepsilon}_{i}\cap\mathbbmss{P}^{\varepsilon}_{i+1}\cap\mathbbmss{R}(\pi_{\varepsilon})\neq\varnothing$. In this case, the local target region is specified as $\mathscr{T}_{i}=(\mathbbmss{P}^{\varepsilon}_{i}\cap\mathbbmss{P}^{\varepsilon}_{i+1}\cap\mathbbmss{R}(\pi_{\varepsilon}))\setminus\mathbf{O}$, and accordingly, the local initial region can be specified as $(\mathbbmss{P}^{\varepsilon}_{i}\cap\mathbbmss{P}^{\varepsilon}_{i-1}\cap\mathbbmss{R}(\pi_{\varepsilon}))\setminus\mathbf{O}$. Based on these three types of local regions, the local LTL formula of each $\mathbbmss{P}_{i}\in\mathbbmss{P}$, which is denoted as $\overbar{\varphi}^{\varepsilon}_{i}$, can be derived as follows.

Let the accepting path be $\pi_{\varepsilon}$. For any pair $(\pi^{\varepsilon}_{i}, \pi^{\varepsilon}_{i+1})$ with $\pi^{\varepsilon}_{i}, \pi^{\varepsilon}_{i+1}\in\pi_{\varepsilon}$ and $i\in\mathbb{N}$, its sub-path is denoted as $\pmb{\mathbbmss{p}}_{i}\subseteq\pmb{\mathbbmss{p}}$, which results in the subsequence $\mathbbmss{P}(\pi^{\varepsilon}_{i}, \pi^{\varepsilon}_{i+1})\subseteq\mathbbmss{P}$. Define $\mathsf{L}_{i}:=|\mathbbmss{P}(\pi^{\varepsilon}_{i}, \pi^{\varepsilon}_{i+1})|$. For each cell $\mathbbmss{P}_{ij}\in\mathbbmss{P}(\pi^{\varepsilon}_{i}, \pi^{\varepsilon}_{i+1})$, where $j\in\mathbbmss{L}_{i}:=\{1, \ldots, \mathsf{L}_{i}\}$, we denote by $\mathscr{T}_{ij}$ the local target region and by $\{\mathscr{I}_{ijl}: l\in\mathbb{L}_{ij}\}$ all local internal regions. The goal in the cell $\mathbbmss{P}_{ij}$ has the following three parts.
\begin{enumerate}[(a)]
  \item If the system state moves into $\mathbbmss{P}^{\varepsilon}_{ij}$, then the system state is expected to stay in $\mathbbmss{P}^{\varepsilon}_{ij}$ while avoiding $\mathbf{O}$. We formulate this part into an LTL formula $\overbar{\varphi}^{\varepsilon}_{ij1}:=\square\tilde{\varphi}^{\varepsilon}_{ij1}$ with $\mathbbmss{R}(\tilde{\varphi}^{\varepsilon}_{ij1})=\mathbbmss{P}^{\varepsilon}_{ij}\setminus\mathbf{O}$.

  \item If the local target region $\mathscr{T}_{ij}$ exists, then in $\mathbbmss{P}^{\varepsilon}_{ij}$ the system state is expected to move into $\mathscr{T}_{ij}$ eventually. This part is formulated into an LTL formula $\overbar{\varphi}^{\varepsilon}_{ij2}:=\diamondsuit\tilde{\varphi}^{\varepsilon}_{ij2}$ with $\mathbbmss{R}(\tilde{\varphi}^{\varepsilon}_{ij2})=\mathscr{T}_{ij}$.

  \item If there exist local internal regions in $\mathbbmss{P}^{\varepsilon}_{ij}$, then the system state is expected to visit all or some of these local internal regions. The following two cases are addressed.
  \begin{itemize}
    \item If $\mathsf{L}_{i}>1$, then only the first and last cells may have local internal regions, i.e, $\mathbbmss{R}(\pi^{\varepsilon}_{i})$ or $\mathbbmss{R}(\pi^{\varepsilon}_{i+1})$. In this case, an LTL formula in $\mathbbmss{P}^{\varepsilon}_{ij}$ (if it exists) is formulated as $\overbar{\varphi}^{\varepsilon}_{ij3}:=\diamondsuit\tilde{\varphi}^{\varepsilon}_{ij3}$, where $\mathbbmss{R}(\tilde{\varphi}^{\varepsilon}_{ij3})\in\{\mathscr{I}_{ijl}: l\in\mathbb{L}_{ij}\}$. In particular, in the last cell of $\mathbbmss{P}(\pi^{\varepsilon}_{i}, \pi^{\varepsilon}_{i+1})$, $\overbar{\varphi}^{\varepsilon}_{ij3}$ can be either $\diamondsuit\tilde{\varphi}^{\varepsilon}_{ij3}$ or $\square\tilde{\varphi}^{\varepsilon}_{ij3}$, which to be chosen depends on the LTL formula $\varphi_{\varepsilon}$. If $(\pi^{\varepsilon}_{i}, \pi^{\varepsilon}_{i+1})$ is the last pair in $\pi_{\varepsilon}$ and the system state is expected to stay in $\mathbbmss{R}(\pi^{\varepsilon}_{i+1})$, then $\overbar{\varphi}^{\varepsilon}_{ij3}=\square\tilde{\varphi}^{\varepsilon}_{ij3}$; otherwise, $\overbar{\varphi}^{\varepsilon}_{ij3}=\diamondsuit\tilde{\varphi}^{\varepsilon}_{ij3}$.

    \item  If $\mathsf{L}_{i}=1$, then the LTL formula $\overbar{\varphi}^{\varepsilon}_{ij3}$ is either $\diamondsuit(\hat{\varphi}^{\varepsilon}_{ij3}\wedge\diamondsuit\check{\varphi}^{\varepsilon}_{ij3})$ or $\diamondsuit(\hat{\varphi}^{\varepsilon}_{ij3}\wedge\square\check{\varphi}^{\varepsilon}_{ij3})$, where $\mathbbmss{R}(\hat{\varphi}^{\varepsilon}_{ij3}), \mathbbmss{R}(\check{\varphi}^{\varepsilon}_{ij3})\in\{\mathscr{I}_{ijl}: l\in\mathbb{L}_{ij}\}$. Similar to the case $\mathsf{L}_{i}>1$, the LTL formula $\varphi_{\varepsilon}$ determines the exact form of $\overbar{\varphi}^{\varepsilon}_{ij3}$. For instance, if $(\pi^{\varepsilon}_{i}, \pi^{\varepsilon}_{i+1})$ is the last pair in $\pi_{\varepsilon}$ and the system state is expected to stay in $\mathbbmss{R}(\pi^{\varepsilon}_{i+1})$, then $\overbar{\varphi}^{\varepsilon}_{ij3}=\diamondsuit(\hat{\varphi}^{\varepsilon}_{ij3}\wedge\square\check{\varphi}^{\varepsilon}_{ij3})$.
  \end{itemize}
\end{enumerate}
Based on the above three parts, we have $\overbar{\varphi}^{\varepsilon}_{ij}:=\overbar{\varphi}^{\varepsilon}_{ij1}\wedge\overbar{\varphi}^{\varepsilon}_{ij2}\wedge\overbar{\varphi}^{\varepsilon}_{ij3}$. From the above three types of local regions, we can see that $\overbar{\varphi}^{\varepsilon}_{i1}$ always exists, whereas the existence of $\overbar{\varphi}^{\varepsilon}_{i2}$ and $\overbar{\varphi}^{\varepsilon}_{i3}$ depends on the existence of local target and internal regions. If all expansion and contraction operators in the above analysis are removed, then we can follow the same mechanism to derive the local LTL formula $\overbar{\varphi}_{ij}$ for the cell $\mathbbmss{P}_{ij}$.

Theorem \ref{thm-1} shows the realization of the pair $(\pi^{\varepsilon}_{i}, \pi^{\varepsilon}_{i+1})$ in $\pi_{\varepsilon}$ via the sequence $\mathbbmss{P}(\pi^{\varepsilon}_{i}, \pi^{\varepsilon}_{i+1})$. To ensure the considered system moves from $\mathbbmss{R}(\pi^{\varepsilon}_{i})$ to $\mathbbmss{R}(\pi^{\varepsilon}_{i+1})$, the considered system needs to satisfy all local LTL formulas $\overbar{\varphi}^{\varepsilon}_{ij}$ sequentially, which results in the following local LTL formula for $\mathbbmss{P}(\pi^{\varepsilon}_{i}, \pi^{\varepsilon}_{i+1})$:
\begin{align}
\label{eqn-13}
\overbar{\varphi}^{\varepsilon}(\pmb{\mathbbmss{p}}_{i}):=\diamondsuit(\overbar{\varphi}^{\varepsilon}_{i1}\wedge\diamondsuit(\overbar{\varphi}^{\varepsilon}_{i2}\wedge\ldots\wedge\diamondsuit\overbar{\varphi}^{\varepsilon}_{i\mathsf{L}_{i}})),
\end{align}
where, the order of all local LTL formulas $\overbar{\varphi}^{\varepsilon}_{ij}$ is based on the sub-path $\pmb{\mathbbmss{p}}_{i}$. For any two successive sequences $\mathbbmss{P}(\pi^{\varepsilon}_{i}, \pi^{\varepsilon}_{i+1})$ and $\mathbbmss{P}(\pi^{\varepsilon}_{i+1}, \pi^{\varepsilon}_{i+2})$, we notice that the last cell in $\mathbbmss{P}(\pi^{\varepsilon}_{i}, \pi^{\varepsilon}_{i+1})$ is the first cell in $\mathbbmss{P}(\pi^{\varepsilon}_{i+1}, \pi^{\varepsilon}_{i+2})$. In this way, the local LTL formula \eqref{eqn-13} can be constructed iteratively. 

In addition, if there exists a loop in $\pi_{\varepsilon}$, then we can consider the following two cases. The first case is that the loop appears only one time. In this case, the loop can be divided into two parts, and a local LTL formula in the form \eqref{eqn-13} can be derived for each part. For instance, if there exists a loop between $\pi^{\varepsilon}_{i}$ and $\pi^{\varepsilon}_{i+1}$, then the local LTL formulas $\overbar{\varphi}^{\varepsilon}(\pmb{\mathbbmss{p}}_{i})$ for $\mathbbmss{P}(\pi^{\varepsilon}_{i}, \pi^{\varepsilon}_{i+1})$ and $\overbar{\varphi}^{\varepsilon}(\pmb{\mathbbmss{p}}_{i+1})$ for $\mathbbmss{P}(\pi^{\varepsilon}_{i+1}, \pi^{\varepsilon}_{i})$ can be derived, and for this loop, the local LTL formula is defined as $\diamondsuit(\overbar{\varphi}^{\varepsilon}(\pmb{\mathbbmss{p}}_{i})\wedge\diamondsuit\overbar{\varphi}^{\varepsilon}(\pmb{\mathbbmss{p}}_{i+1}))$. The second case is that the loop appears infinitely. Similar to the first case, the loop is divided into two parts and the local LTL formulas for these two parts can be derived. The the local LTL formula for this loop is defined as $\square(\diamondsuit(\overbar{\varphi}^{\varepsilon}(\pmb{\mathbbmss{p}}_{i})\wedge\diamondsuit\overbar{\varphi}^{\varepsilon}(\pmb{\mathbbmss{p}}_{i+1})))$. From the above analysis and the path $\pmb{\mathbbmss{p}}$, the local LTL formulas for all $\mathbbmss{P}_{i}\in\mathbbmss{P}$ can be combined into the following form:
\begin{align}
\label{eqn-14}
&\diamondsuit(\overbar{\varphi}^{\varepsilon}_{0}\wedge\diamondsuit(\overbar{\varphi}^{\varepsilon}_{1}\wedge\ldots\wedge\square(\diamondsuit(\overbar{\varphi}^{\varepsilon}_{j}\wedge
\diamondsuit(\overbar{\varphi}^{\varepsilon}_{j+1}\wedge\ldots\wedge\diamondsuit\overbar{\varphi}^{\varepsilon}_{j+k}))) \nonumber \\
&\wedge\ldots\wedge\diamondsuit\overbar{\varphi}^{\varepsilon}_{\mathsf{L}})),
\end{align}
where, $\overbar{\varphi}^{\varepsilon}_{i}$ is the local LTL formula corresponding to each cell $\mathbbmss{P}_{i}\in\mathbbmss{P}$, $i\in\mathbbmss{L}=\{0, \ldots, \mathsf{L}\}$ and $\mathsf{L}=|\mathbbmss{P}|$. Note that \eqref{eqn-14} is based on the robust accepting path $\pi_{\varepsilon}$. For the accepting path $\pi$, we can follow the above mechanism to establish all local LTL formulas, which are denoted as $\overbar{\varphi}_{i}$ and whose combination is of the following form:
\begin{align}
\label{eqn-15}
&\diamondsuit(\overbar{\varphi}_{0}\wedge\diamondsuit(\overbar{\varphi}_{1}\wedge\ldots\wedge\square(\diamondsuit(\overbar{\varphi}_{j}\wedge\diamondsuit(\overbar{\varphi}_{j+1}\wedge\ldots
\wedge\diamondsuit\overbar{\varphi}_{j+k})))  \nonumber \\
&\wedge\ldots\wedge\diamondsuit\overbar{\varphi}_{\mathsf{L}})).
\end{align}
Each $\varphi_{i}$ corresponds to $\overbar{\varphi}^{\varepsilon}_{i}$. Furthermore, the following proposition is derived directly.

\begin{proposition}
\label{prop-1}
Consider the LTL formula $\varphi$ and $\varepsilon>0$. Let the decompositions \eqref{eqn-14} and \eqref{eqn-15} be derived as above. $\varphi$ is satisfied if all $\overbar{\varphi}_{i}$ in \eqref{eqn-14} are satisfied. Each $\varphi_{i}$ is satisfied if the corresponding $\overbar{\varphi}^{\varepsilon}_{i}$ are satisfied.
\end{proposition}

With Proposition \ref{prop-1}, the controller synthesis problem (i.e., \textbf{Problem 2}) is decomposed into finite local ones, which will be discussed in detail in the next section.

\section{Abstraction-based Controller Synthesis}
\label{sec-abstractionbased}

After solving \textbf{Problem 1}, we propose an abstraction-based local-to-global control strategy to deal with \textbf{Problem 2} in this section. To be specific, we take advantage of the properties of zonotopes to propose a novel approximation of each cell in $\mathbbmss{P}$ in Section \ref{subsec-approximate} and to construct the local symbolic abstraction in Section \ref{subsec-construction}. The local-to-global abstract control strategy is proposed in Section \ref{subsec-consyn}.

\subsection{Approximation of State and Input Sets}
\label{subsec-approximate}

Note that each cell in $\mathbbmss{P}$ is either a zonotope or a constrained zonotope. Without loss of generality, we focus on the approximation of any zonotope $\mathbf{Z}\subset X_{1}$ and constrained zonotope $\mathbf{Z}^{\cstr}\subset X_{1}$ in this subsection.

\begin{figure}[!t]
\begin{center}
\begin{picture}(70, 88)
\put(-64, -15){\resizebox{70mm}{35mm}{\includegraphics[width=2.5in]{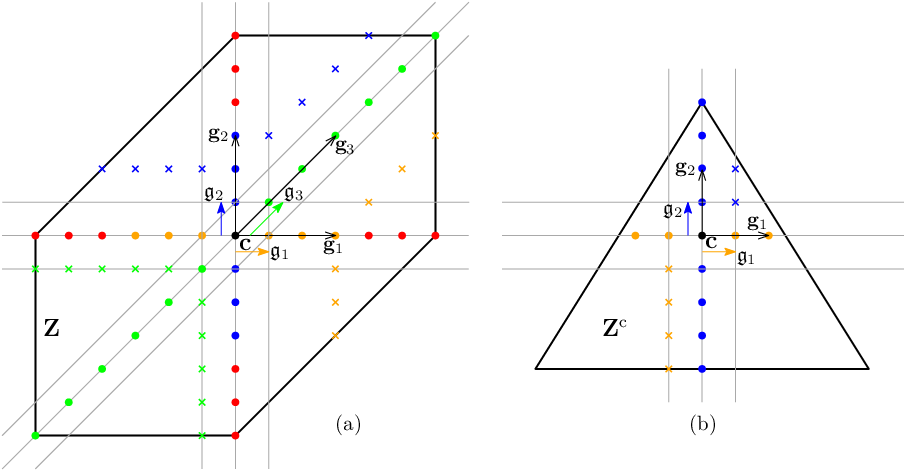}}}
\end{picture}
\end{center}
\caption{Illustration of the proposed approximation approach. (a) Approximation of the zonotope $\mathbf{Z}=\{\mathbf{c}, \mathbf{G}\}\subset\mathbb{R}^{2}$ with $\mathbf{G}=(\mathbf{g}_{1}, \mathbf{g}_{2}, \mathbf{g}_{3})$. (b) Approximation of the constrained zonotope $\mathbf{Z}^{\cstr}=\{\mathbf{c}, \mathbf{G}, \mathbf{A}, \mathbf{b}\}\subset\mathbb{R}^{2}$ with $\mathbf{G}=(\mathbf{g}_{1}, \mathbf{g}_{2})$. The dots belong to the set $\mathbbmss{F}_{l}$ or $\mathbbmss{F}^{\cstr}_{l}$, whereas the crosses belong to the set $\overbar{\mathbbmss{F}}_{l}$ or $\overbar{\mathbbmss{F}}^{\cstr}_{l}$.}
\label{fig-6}
\end{figure}

\subsubsection{Approximation of Zonotopes}
Different from the classic quantization-based technique \cite{Pola2008approximately}, we approximate the zonotope $\mathbf{Z}\subset X_{1}$ via the properties of zonotopes. Let $\mathbf{Z}=\{\mathbf{c}, \mathbf{G}\}$ with $\mathbf{G}=(\mathbf{g}_{1}, \ldots, \mathbf{g}_{\ell})$ and $\pmb{\ell}:=\{1, \ldots, \ell\}$, where $\ell\in\mathbb{N}$.

First, each $\mathbf{g}_{l}$ with $l\in\pmb{\ell}$ is partitioned uniformly. Let $\mathsf{N}_{l}\in\mathbb{N}$ be the partition number, and define $\mathbf{G}_{\mathsf{b}}:=(\mathfrak{g}_{1}, \ldots, \mathfrak{g}_{\ell})=(\mathsf{N}^{-1}_{1}\mathbf{g}_{1}, \ldots, \mathsf{N}^{-1}_{\ell}\mathbf{g}_{\ell})$ as the basic generator matrix. Thus, $(\mathbf{g}_{1}, \ldots, \mathbf{g}_{\ell})=(\mathsf{N}_{1}\mathfrak{g}_{1}, \ldots, \mathsf{N}_{\ell}\mathfrak{g}_{\ell})$. Since $\mathfrak{g}_{l}$ is the basis of $\mathbf{g}_{l}$, we derive a finite number of points consisting of the set below
\begin{align}
\label{eqn-16}
\mathbbmss{F}_{l}:=\{\mathbf{c}\pm\mathfrak{g}_{l}, \ldots, \mathbf{c}\pm\mathsf{N}_{l}\mathfrak{g}_{l}, \ldots, \mathbf{c}\pm\mathsf{M}_{l}\mathfrak{g}_{l}\},
\end{align}
where $\mathsf{M}_{l}\geq\mathsf{N}_{l}$ is the largest integer such that $\mathbf{c}\pm\mathsf{M}_{l}\mathfrak{g}_{l}\in\mathbf{Z}$; see the red points in Fig. \ref{fig-6}(a). In \eqref{eqn-16}, $\mathsf{M}_{l}$ depends on other generators $\{\mathfrak{g}_{1}, \ldots, \mathfrak{g}_{l-1}, \mathfrak{g}_{l+1}, \ldots, \mathfrak{g}_{\ell}\}$, and $\mathsf{N}_{l}$ is constrained by a parameter $\mu>0$. That is, $\max_{l\in\pmb{\ell}}\{\|\mathfrak{g}_{l}\|\}\leq\mu$ is imposed to constrain the choice of $\{\mathsf{N}_{l}: l\in\pmb{\ell}\}$. $\mu>0$ depends on the desired equivalence relation; see Section \ref{subsec-construction}. In addition, the choice of the set $\{\mathsf{N}_{l}: l\in\pmb{\ell}\}$ affects the set $\{\mathsf{M}_{l}: l\in\pmb{\ell}\}$.

Second, all points in $\mathbbmss{F}_{l}$ are the bases to generate other points in $\mathbf{Z}$. Given a $\mathsf{p}_{lj}\in\mathbbmss{F}_{l}$, $0<j\leq\mathsf{M}_{l}$, other points are generated in the following way: we first take $\mathsf{p}_{lj}\in\mathbbmss{F}_{l}$ as the basis point and any generator from $\{\mathfrak{g}_{1}, \ldots, \mathfrak{g}_{\ell}\}$ as the direction, and then generate all the points leading to the following set
\begin{align}
\label{eqn-17}
\overbar{\mathbbmss{F}}_{l}:=\bigcup_{0<j\leq\overbar{\mathsf{N}}_{l}}\bigcup_{k\in\pmb{\ell}, k\neq l}\{\mathsf{p}_{lj}\pm\mathfrak{g}_{k}, \ldots, \mathsf{p}_{lj}\pm\mathsf{N}_{k}\mathfrak{g}_{k}\}\cap\mathbf{Z}.
\end{align}
This generation mechanism is terminated until the generated point is not in $\mathbf{Z}$. Since $\mathbbmss{F}_{l}$ is finite, this generation mechanism can be implemented recursively and terminated in finite steps.

Finally, combining all generated points, the zonotope $\mathbf{Z}$ is approximated by the following set:
\begin{align}
\label{eqn-18}
\mathscr{A}(\mathbf{Z}):=\{\mathbf{c}\}\cup(\cup_{l\in\pmb{\ell}}(\mathbbmss{F}_{l}\cup\overbar{\mathbbmss{F}}_{l})).
\end{align}

\begin{remark}
\label{rmk-4}
We examine the above approximation in detail, and compare the proposed approach with the existing ones.
\begin{enumerate}[(i)]
  \item The proposed approach is based on the properties of zonotopes. That is, the center and all generators of $\mathbf{Z}$ are applied in the approximation, which implies that each zonotope can have its own approximation. However, the classic quantization-based techniques \cite{Pola2008approximately, Meyer2019hierarchical, Ren2019dynamic} are to approximate the whole state space uniformly.

  \item The proposed approach can be simplified by choosing $n$ linearly independent generators from $\{\mathbf{g}_{1}, \ldots, \mathbf{g}_{\ell}\}$. Note that $\mathbf{G}=(\mathbf{g}_{1}, \ldots, \mathbf{g}_{\ell})$ is full-rank and $\ell\geq n$. Hence, we can choose $n$ linear independent generators $\{\mathbf{g}_{l_{1}}, \ldots, \mathbf{g}_{l_{n}}\}\subseteq\{\mathbf{g}_{1}, \ldots, \mathbf{g}_{\ell}\}$ with $\{l_{1}, \ldots, l_{n}\}\subseteq\pmb{\ell}$ and derive a novel basic generator matrix $\mathbf{G}'_{\textrm{b}}$. The following approximation can be proceeded similarly. For instance, we can choose any two generators in Fig. \ref{fig-6}(a). This simplification reduces the number of abstract states in \eqref{eqn-18}, and will potentially reduce the transition number in the symbolic abstraction and further affect the computational complexity. If all generators are orthogonal, then the proposed approach is reduced to those in many existing works \cite{Zamani2015symbolic, Girard2007approximation}. In this aspect, the proposed approximation approach is more general.

  \item Each generator has its own partition number and hence has its own partition. Here we present two special cases to unify the partition of all generators. The first case is to select basic generators with the same length for all $\{\mathbf{g}_{1}, \ldots, \mathbf{g}_{\ell}\}$ such that the partition is unified. Let the length be $\mathsf{h}>0$. Hence, $\mathfrak{g}_{l}:=\mathsf{h}\mathbf{g}_{l}/\|\mathbf{g}_{l}\|$, $\mathsf{N}_{l}:=\lfloor\|\mathbf{g}_{l}\|/\|\mathfrak{g}_{l}\|\rfloor$ with the ceiling operator $\lfloor\cdot\rfloor$, and the set $\mathbbmss{F}_{l}$ in \eqref{eqn-16} can be defined similarly. The second case is to set the same partition number for all $\{\mathbf{g}_{1}, \ldots, \mathbf{g}_{\ell}\}$. That is, $\mathsf{N}_{l}=\mathsf{N}\in\mathbb{N}$ for all $l\in\pmb{\ell}$. Hence, the set $\mathbbmss{F}_{l}$ in \eqref{eqn-16} can be defined similarly. After defining the set $\mathbbmss{F}_{l}$, the following approximation is proceeded similarly.
      \hfill $\vartriangle$
\end{enumerate}
\end{remark}

\subsubsection{Approximation of Constrained Zonotopes}
Following the above techniques, any constrained zonotope $\mathbf{Z}^{\cstr}=\{\mathbf{c}, \mathbf{G}, \mathbf{A}, \mathbf{b}\}$ can be approximated similarly. The only difference lies in the constraint condition $\mathbf{A}\xi=\mathbf{b}$, which has effects on $\mathsf{M}_{l}$ and further on the set $\mathbbmss{F}_{l}$ as in \eqref{eqn-16}. In addition, the set $\overbar{\mathbbmss{F}}_{l}$ in \eqref{eqn-17} is modified to be
\begin{align}
\label{eqn-19}
\overbar{\mathbbmss{F}}^{\cstr}_{l}&:=\bigcup_{0<j\leq\overbar{\mathsf{N}}_{l}}\bigcup_{k\in\pmb{\ell}, k\neq l}\{\mathsf{p}_{lj}\pm\mathfrak{g}_{k}, \ldots, \mathsf{p}_{lj}\pm\mathsf{M}
_{k}\mathfrak{g}_{k}\}\cap\mathbf{Z}^{\cstr}.
\end{align}
Therefore, $\mathbf{Z}^{\cstr}$ is approximated via the following set
\begin{align}
\label{eqn-20}
\mathscr{A}(\mathbf{Z}^{\cstr}):=\{\mathbf{c}\}\cup(\cup_{l\in\pmb{\ell}}(\mathbbmss{F}^{\cstr}_{l}\cup\overbar{\mathbbmss{F}}^{\cstr}_{l})).
\end{align}

\begin{remark}
\label{rmk-5}
Besides the proposed approach, other approaches like tiling and zonotopal subdivision techniques \cite{Mcmullen1975space, Kabi2020synthesizing} can be applied to the approximation of zonotopes. These techniques take centrally symmetric advantages of zonotopes to partition zonotopes into sub-zonotopes. For instance, the connection of all generated points leads to a line arrangement, which determines the zonotope division \cite{Richter1993line}. However, these techniques may result in huge constructional and computational complexities. How to apply these techniques to approximate constrained zonotopes is unknown, since constrained zonotopes are not necessarily centrally symmetric.
\hfill $\vartriangle$
\end{remark}

With the approximation of (constrained) zonotopes, we introduce the norm $\|\cdot\|_{\mathbf{G}}$ based on the generator matrix $\mathbf{G}$ and the vector projection. Given any $\mathbf{v}\in\mathbb{R}^{n}$, $\|\mathbf{v}\|_{\mathbf{G}}:=\max_{l\in\pmb{\ell}}\{\mathbf{v}\cdot\mathbf{g}_{l}/\|\mathbf{g}_{l}\|\}$, where $\mathbf{v}\cdot\mathbf{g}_{l}/\|\mathbf{g}_{l}\|$ is the scalar projection of $\mathbf{v}$ onto $\mathbf{g}_{l}$. If all generators are orthogonal, then this norm is reduced to the infinity norm. With the norm $\|\cdot\|_{\mathbf{G}}$, we can see that for any $x\in\mathbf{Z}$ (or $x\in\mathbf{Z}^{\cstr}$), there exists $q\in\mathscr{A}(\mathbf{Z})$ (or $q\in\mathscr{A}(\mathbf{Z}^{\cstr})$) such that $\|x-q\|_{\mathbf{G}}\leq0.5\max_{l\in\pmb{\ell}}\{\|\mathfrak{g}_{l}\|\}$.

\subsubsection{Approximation of Input Set}
For a zonotope $\mathbf{Z}\subset X_{1}$, its input set is defined as $U_{1}(\mathbf{Z}):=\cup_{x\in\mathbf{Z}}\enab(x)\subseteq U_{1}$, and approximated as follows. Given any $q\in\mathscr{A}(\mathbf{Z})$, the reachable set of $\mathbf{T}_{\tau}(\Sigma, \mathbf{Z})$ from $q$ is $\reach(\tau, q):=\{x'\in\mathbf{Z}: x'=\mathbf{x}(\tau, q, u), u\in U_{1}(\mathbf{Z})\}$, which is well-defined due to $U_{1}(\mathbf{Z})$. Given any $\eta\in\mathbb{R}^{+}$, consider the set $\mathbbmss{S}_{\eta}(\tau, q):=\{v\in\mathscr{A}(\mathbf{Z}):$ there exists $z\in\reach(\tau, q)$ such that $\|v-z\|_{\mathbf{G}}\leq0.5\eta\}$, which is a countable set. Note that $\eta$ is constrained by the desired precision; see Section \ref{subsec-construction}. A function $\psi: \mathbbmss{S}_{\eta}(\tau, q)\rightarrow U_{1}(\mathbf{Z})$ is defined: for any $v\in\mathbbmss{S}_{\eta}(\tau, q)$, there exists $u_{1}=\psi(v)\in U_{1}(\mathbf{Z})$ such that $\|v-\mathbf{x}(\tau, q, u_{1})\|_{\mathbf{G}}\leq0.5\eta$. Let $U_{2}(q):=\psi(\mathbbmss{S}_{\eta}(\tau, q))$, which captures all inputs that can be applied at $q\in\mathscr{A}(\mathbf{Z})$. The set $U_{2}(q)$ is countable since $U_{2}(q)$ is the image of the function $\psi$ on $\mathbbmss{S}_{\eta}(\tau, q)$. Hence, $U_{1}(\mathbf{Z})$ is approximated via the following countable set:
\begin{equation}
\label{eqn-21}
U_{2}(\mathbf{Z}):=\bigcup\nolimits_{q\in\mathscr{A}(\mathbf{Z})}U_{2}(q).
\end{equation}
That is, the set $U_{2}(\mathbf{Z})$ approximates the set $U_{1}(\mathbf{Z})$ in the following way: given any $q\in\mathscr{A}(\mathbf{Z})$ and any $u_{1}\in U_{1}(\mathbf{Z})$, there exists $u_{2}\in U_{2}(q)$ such that $\|\mathbf{x}(\tau, q, u_{1})-\mathbf{x}(\tau, q, u_{2})\|_{\mathbf{G}}\leq\eta$.

The above approximation can be applied similarly to constrained zonotopes, which is omitted here. From \eqref{eqn-4} and \eqref{eqn-11}, all cells in $\mathbb{Z}$ are (constrained) zonotopes, and hence the above approximation can also be applied to any cell in $\mathbb{Z}$.

\subsection{Abstraction Construction}
\label{subsec-construction}

With the approximation in the previous subsection, the symbolic abstraction is constructed in this subsection. For each cell $\mathbbmss{P}_{i}\in\mathbbmss{P}$, $i\in\mathbbmss{L}$, the symbolic abstraction for the system $\mathbf{T}_{\tau}(\Sigma, \mathbbmss{P}_{i})$ is constructed as a transition system $\mathbf{T}_{\tau, \mu_{i}, \eta_{i}}(\Sigma, \mathbbmss{P}_{i})=(X_{2i}, X^{0}_{2i}, U_{2i}, \Delta_{2i}, Y_{2i}, H_{2i})$, where,
\begin{itemize}
\item the set of states is $X_{2i}=\mathscr{A}(\mathbbmss{P}_{i})$;
\item the set of initial states is $X^{0}_{2i}=\mathscr{A}(\mathbbmss{P}^{i}_{0})$ with $\mathbbmss{P}^{i}_{0}\subseteq\mathbbmss{P}_{i}$;
\item the set of inputs is $U_{2i}=U_{2}(\mathbbmss{P}_{i})$;
\item the transition relation is given as follows: for $q_{1}, q_{2}\in X_{2i}$ and $u\in U_{2i}$, $q_{2}\in\Delta_{2i}(q_{1}, u)$ if and only if
\begin{align}
\label{eqn-22}
\hspace{-10pt} q_{2}\in\{\bar{q}\in X_{2i}: \|\mathbf{x}(\tau, q_{1}, u)-\bar{q}\|_{\mathbf{G}}\leq(1+e^{L_{i}\tau})\varepsilon\}, 
\end{align}
where $L_{i}>0$ is the local Lipschitz constant of $f$ in $\mathbbmss{P}_{i}$; 
\item the set of outputs is $Y_{2i}=X_{2i}$;
\item the output map is $H_{2i}=\Id$.
\end{itemize}
The parameters $\mu_{i}, \eta_{i}>0$ are from the approximation of $\mathbbmss{P}_{i}\in\mathbbmss{P}$. $\mathbf{T}_{\tau, \mu_{i}, \eta_{i}}(\Sigma, \mathbbmss{P}_{i})$ is metric when $Y_{2i}$ is equipped with the metric $\mathbf{d}(y', y)=\|y-y'\|$ for all $y, y'\in Y_{2i}$. $\mathbf{T}_{\tau, \mu_{i}, \eta_{i}}(\Sigma, \mathbbmss{P}_{i})$ is nondeterministic from the transition relation \eqref{eqn-22}. Next, we show the FRR from $\mathbf{T}_{\tau}(\Sigma, \mathbbmss{P}_{i})$ to $\mathbf{T}_{\tau, \mu_{i}, \eta_{i}}(\Sigma, \mathbbmss{P}_{i})$.

\begin{theorem}
\label{thm-2}
Consider the systems $\mathbf{T}_{\tau, \mu_{i}, \eta_{i}}(\Sigma, \mathbbmss{P}_{i})$ and $\mathbf{T}_{\tau}(\Sigma, \mathbbmss{P}_{i})$ with the parameters $\tau, \mu_{i}, \eta_{i}\in\mathbb{R}^{+}$. Given a precision $\varepsilon\in\mathbb{R}^{+}$, if the relation $\mathscr{F}: X_{1i}\rightarrow X_{2i}$ is defined as
\begin{equation}
\label{eqn-23}
\mathscr{F}(x)=\{q\in X_{2i}: \|x-q\|_{\mathbf{G}}\leq\varepsilon, x\in X_{1i}\},
\end{equation}
then $\mathbf{T}_{\tau, \mu_{i}, \eta_{i}}(\Sigma, \mathbbmss{P}_{i})\preceq_{\mathscr{F}}\mathbf{T}_{\tau}(\Sigma, \mathbbmss{P}_{i})$.
\end{theorem}

\begin{IEEEproof}
See Appendix \ref{prf-thm2}.
\end{IEEEproof}

Since the state space is constrained into $\mathbbmss{P}_{i}\in\mathbbmss{P}$, the symbolic abstraction is local and can be different for different cells in $\mathbbmss{P}$, which allows for the flexibility in the abstraction construction. If the system $\Sigma$ satisfies the stability condition as in \cite{Girard2012controller}, then a special case of the symbolic abstraction can be derived. To be specific, the system $\Sigma$ is \emph{incrementally globally asymptotically stable ($\delta$-GAS)} in $\mathbbmss{P}_{i}$, if there exists $\beta_{i}\in\mathcal{KL}$ such that for all $x_{1}, x_{2}\in\mathbbmss{P}_{i}, u\in{U}_{1}(\mathbbmss{P}_{i})$ and $t\in\mathbb{R}^{+}$,
\begin{align}
\label{eqn-24}
\|\mathbf{x}(t, x_{1}, u)-\mathbf{x}(t, x_{2}, u)\|_{\mathbf{G}}\leq\beta_{i}(\|x_{1}-x_{2}\|_{\mathbf{G}}, t).
\end{align}
Note that the proposed norm in Section \ref{subsec-approximate} does not affect the incremental stability due to the norm equivalence. In this case, the symbolic abstraction of $\mathbf{T}_{\tau}(\Sigma, \mathbbmss{P}_{i})$ is constructed as $\overbar{\mathbf{T}}_{\tau, \mu_{i}, \eta_{i}}(\Sigma, \mathbbmss{P}_{i})=(\bar{X}_{2i}, \bar{X}^{0}_{2i}, \bar{U}_{2i}, \bar{\Delta}_{2i}, \bar{Y}_{2i}, \bar{H}_{2i})$,
\begin{itemize}
\item the set of states is $\bar{X}_{2i}=\mathscr{A}(\mathbbmss{P}_{i})$;
\item the set of initial states is $\bar{X}^{0}_{2i}=\mathscr{A}(\mathbbmss{P}^{i}_{0})$ with $\mathbbmss{P}^{i}_{0}\subseteq\mathbbmss{P}_{i}$;
\item the set of inputs is $\bar{U}_{2i}=U_{2}(\mathbbmss{P}_{i})$;
\item the transition relation is given as follows: for $q_{1}, q_{2}\in\bar{X}_{2i}$ and $u\in\bar{U}_{2i}$, $q_{2}\in\bar{\Delta}_{2i}(q_{1}, u)$ if and only if
\begin{align}
\label{eqn-25}
q_{2}\in\{\bar{q}\in\bar{X}_{2i}: \|\mathbf{x}(\tau, q_{1}, u)-\bar{q}\|_{\mathbf{G}}\leq0.5\mu_{i}\};
\end{align}
\item the set of outputs is $\bar{Y}_{2i}=\bar{X}_{2i}$;
\item the output map is $\bar{H}_{2i}=\Id$.
\end{itemize}

Similar to Theorem \ref{thm-2}, the following theorem establishes the $\varepsilon$-ABR between $\overbar{\mathbf{T}}_{\tau, \mu_{i}, \eta_{i}}(\Sigma, \mathbbmss{P}_{i})$ and $\mathbf{T}_{\tau}(\Sigma, \mathbbmss{P}_{i})$.

\begin{theorem}
\label{thm-3}
Consider the $\delta$-GAS system $\mathbf{T}_{\tau}(\Sigma, \mathbbmss{P}_{i})$ and its symbolic model $\overbar{\mathbf{T}}_{\tau, \mu_{i}, \eta_{i}}(\Sigma, \mathbbmss{P}_{i})$ with the parameters $\tau, \mu_{i}, \eta_{i}\in\mathbb{R}^{+}$. Given a precision $\varepsilon>0$, if 
\begin{align}
\label{eqn-26}
\beta_{i}(\varepsilon, \tau)+\mu_{i}+0.5\eta_{i}\leq\varepsilon,
\end{align}
where $\beta_{i}$ is given in \eqref{eqn-24}, then $\overbar{\mathbf{T}}_{\tau, \mu_{i}, \eta_{i}}(\Sigma, \mathbbmss{P}_{i})\simeq_{\varepsilon}\mathbf{T}_{\tau}(\Sigma, \mathbbmss{P}_{i})$.
\end{theorem}

\begin{IEEEproof}
See Appendix \ref{prf-thm3}.
\end{IEEEproof}

In the construction of $\mathbf{T}_{\tau, \mu_{i}, \eta_{i}}(\Sigma, \mathbbmss{P}_{i})$ and $\overbar{\mathbf{T}}_{\tau, \mu_{i}, \eta_{i}}(\Sigma, \mathbbmss{P}_{i})$, the difference is the transition relation. The transition relation in $\mathbf{T}_{\tau, \mu_{i}, \eta_{i}}(\Sigma, \mathbbmss{P}_{i})$ is non-deterministic and the FRR defines which transition to be chosen, whereas the transition relation in  $\overbar{\mathbf{T}}_{\tau, \mu_{i}, \eta_{i}}(\Sigma, \mathbbmss{P}_{i})$ is deterministic and the $\varepsilon$-ABR constrains the choices of the parameters via \eqref{eqn-26}. From Theorems \ref{thm-2}-\ref{thm-3}, the proposed construction approach can be applied to general nonlinear systems, which does not necessarily need to be $\delta$-GAS. In addition, the global $\delta$-GAS assumption in \cite{Girard2012controller, Pola2008approximately, Girard2010approximately} is relaxed into the local version \eqref{eqn-24}, which implies that the proposed approach is less conservative.

For each cell $\mathbbmss{P}_{i}\in\mathbbmss{P}$, $i\in\mathbbmss{L}$, the symbolic abstraction can be constructed as either $\mathbf{T}_{\tau, \mu_{i}, \eta_{i}}(\Sigma, \mathbbmss{P}_{i})$ or $\overbar{\mathbf{T}}_{\tau, \mu_{i}, \eta_{i}}(\Sigma, \mathbbmss{P}_{i})$. That is, different cells have their own symbolic models such that either the FRR or the $\varepsilon$-ABR is satisfied. To unify the notations for all cells, we use $\mathcal{T}_{\tau, \mu_{i}, \eta_{i}}(\Sigma, \mathbbmss{P}_{i})$ to denote the symbolic abstraction. All parameters $\mu_{i}, \eta_{i}$ with $i\in\mathbbmss{L}$ are combined as $\pmb{\mu}:=(\mu_{0}, \ldots, \mu_{\mathsf{L}})$ and $\pmb{\eta}:=(\eta_{0}, \ldots, \eta_{\mathsf{L}})$. For the set $\mathbbmss{P}$,  its symbolic abstraction is denoted as $\mathcal{T}_{\tau, \pmb{\mu}, \pmb{\eta}}(\Sigma, \mathbbmss{P})$, which is the union of all $\mathcal{T}_{\tau, \mu_{i}, \eta_{i}}(\Sigma, \mathbbmss{P}_{i})$.

\subsection{Local-to-Global Controller Strategy}
\label{subsec-consyn}

Using the abstraction-based approach, the controller strategy is derived in this subsection. For this purpose, we first show the relations between global and local controllers.

\begin{theorem}
\label{thm-4}
Consider the symbolic model $\mathcal{T}_{\tau, \pmb{\mu}, \pmb{\eta}}(\Sigma, \mathbbmss{P})$ and the LTL formula $\varphi$. If for all $i\in\mathbbmss{L}$, there exists a local abstract controller $\mathscr{C}^{i}_{\mathsf{a}}$ such that the local LTL formula $\overbar{\varphi}^{\varepsilon}_{i}$ is satisfied for $\mathcal{T}_{\tau, \mu_{i}, \eta_{i}}(\Sigma, \mathbbmss{P}_{i})$, then all local abstract controllers are combined such that $\varphi$ is satisfied for $\mathcal{T}_{\tau, \pmb{\mu}, \pmb{\eta}}(\Sigma, \mathbbmss{P})$.
\end{theorem}

\begin{IEEEproof}
From Proposition \ref{prop-1}, the satisfaction of the LTL formula $\varphi$ equals to the satisfaction of all local LTL formulas $\overbar{\varphi}^{\varepsilon}_{i}$, $i\in\mathbbmss{L}$. Since the local abstract controller $\mathscr{C}^{i}_{\mathsf{a}}$ ensures the satisfaction of the local LTL formula $\overbar{\varphi}^{\varepsilon}_{i}$ for $\mathcal{T}_{\tau, \mu_{i}, \eta_{i}}(\Sigma, \mathbbmss{P}_{i})$, all local abstract controllers can be combined as the following global abstract controller
\begin{equation}
\label{eqn-27}
\mathscr{C}_{\mathsf{a}}(q):=\mathscr{C}^{i}_{\mathsf{a}}(q), \quad \forall q\in\mathscr{A}(\mathbbmss{P}_{i}), i\in\mathbbmss{L},
\end{equation}
which further guarantees the satisfaction of $\varphi$ for the system $\mathcal{T}_{\tau, \pmb{\mu}, \pmb{\eta}}(\Sigma, \mathbbmss{P})$ and thus completes the proof.
\end{IEEEproof}

Theorem \ref{thm-4} shows how to construct the global abstract controller \eqref{eqn-27} from all local abstract controllers. From Theorem \ref{thm-4}, we can derive the similar result to bridge the controllers for the LTL formulas $\varphi$ and $\overbar{\varphi}_{i}$. That is, if $\mathscr{C}_{i}$ is the local controller such that $\overbar{\varphi}_{i}$ is satisfied for $\mathcal{T}_{\tau}(\Sigma, \mathbbmss{P}_{i})$, then the global controller for  $\varphi$ is constructed as
\begin{equation}
\label{eqn-28}
\mathscr{C}(x):=\mathscr{C}_{i}(x), \quad \forall x\in\mathbbmss{P}_{i}, i\in\mathbbmss{L},
\end{equation}
which shows the relation between global and local controllers. In the following, we establish the relation between the controllers for the LTL formulas $\overbar{\varphi}^{\varepsilon}_{i}$ and $\overbar{\varphi}_{i}$.

\begin{theorem}
\label{thm-5}
Consider the systems $\mathcal{T}_{\tau, \mu_{i}, \eta_{i}}(\Sigma, \mathbbmss{P}_{i})$ and $\mathbf{T}_{\tau}(\Sigma, \mathbbmss{P}_{i})$. Let $\varepsilon>0$ be a given precision and the local LTL formulas be $\overbar{\varphi}_{i}$ and $\overbar{\varphi}^{\varepsilon}_{i}$.
\begin{enumerate}[(1)]
  \item Let $\mathcal{T}_{\tau, \mu_{i}, \eta_{i}}(\Sigma, \mathbbmss{P}_{i})=\mathbf{T}_{\tau, \mu_{i}, \eta_{i}}(\Sigma, \mathbbmss{P}_{i})$ and $\mathbf{T}_{\tau}(\Sigma, \mathbbmss{P})\preceq_{\mathscr{F}}\mathbf{T}_{\tau, \mu_{i}, \eta_{i}}(\Sigma, \mathbbmss{P}_{i})$. If there exists a local abstract controller $\mathscr{C}^{i}_{\mathsf{a}}: \mathscr{A}(\mathbbmss{P}^{\varepsilon}_{i})\rightarrow2^{U_{2}(\mathbbmss{P}^{\varepsilon}_{i})}$ such that $\overbar{\varphi}^{\varepsilon}_{i}$ is satisfied for $\mathbf{T}_{\tau, \mu_{i}, \eta_{i}}(\Sigma, \mathbbmss{P}_{i})$, then there exists a local controller
      \begin{equation}
      \label{eqn-29}
      \mathscr{C}_{i}(x):=\mathscr{C}^{i}_{\mathsf{a}}(\mathscr{F}(x)), \quad x\in\mathbbmss{P}_{i}
      \end{equation}
      such that $\overbar{\varphi}_{i}$ is satisfied for $\mathbf{T}_{\tau}(\Sigma, \mathbbmss{P}_{i})$.

  \item Let $\mathcal{T}_{\tau, \mu_{i}, \eta_{i}}(\Sigma, \mathbbmss{P}_{i})=\overbar{\mathbf{T}}_{\tau, \mu_{i}, \eta_{i}}(\Sigma, \mathbbmss{P}_{i})$ and $\mathbf{T}_{\tau}(\Sigma, \mathbbmss{P})\simeq_{\varepsilon}\overbar{\mathbf{T}}_{\tau, \mu_{i}, \eta_{i}}(\Sigma, \mathbbmss{P}_{i})$ with the $\varepsilon$-ABR $\mathscr{R}$. If there exists a local abstract controller $\mathscr{C}^{i}_{\mathsf{a}}: \mathscr{A}(\mathbbmss{P}^{\varepsilon}_{i})\rightarrow2^{U_{2}(\mathbbmss{P}^{\varepsilon}_{i})}$ such that $\overbar{\varphi}^{\varepsilon}_{i}$ is satisfied for $\overbar{\mathbf{T}}_{\tau, \mu_{i}, \eta_{i}}(\Sigma, \mathbbmss{P}_{i})$, then the local controller
      \begin{equation}
      \label{eqn-30}
      \mathscr{C}_{i}(x):=\mathscr{C}^{i}_{\mathsf{a}}(\mathscr{R}(x)), \quad x\in\mathbbmss{P}_{i}
      \end{equation}
     is such that $\overbar{\varphi}_{i}$ is satisfied for $\mathbf{T}_{\tau}(\Sigma, \mathbbmss{P}_{i})$.
\end{enumerate}
\end{theorem}

\begin{IEEEproof}
Since we focus on the local state space $\mathbbmss{P}_{i}$ instead of the whole state space $\mathbb{X}$, the local abstract controller $\mathscr{C}^{i}_{\mathsf{a}}$ is defined on $\mathscr{A}(\mathbbmss{P}^{\varepsilon}_{i})$ such that the controllers \eqref{eqn-29}-\eqref{eqn-30} are well-defined on $\mathbbmss{P}_{i}$. That is, for any $x\in\mathbbmss{P}_{i}$, $\mathscr{F}(x)\in\mathscr{A}(\mathbbmss{P}^{\varepsilon}_{i})$ holds from the FRR, and $\mathscr{R}(x)\in\mathscr{A}(\mathbbmss{P}^{\varepsilon}_{i})$ holds from the $\varepsilon$-ABR. Next, we can follow \cite[Theorem VI.3]{Reissig2017feedback} and \cite[Theorem 5.1]{Tabuada2008approximate} respectively to show that, under the controllers \eqref{eqn-29}-\eqref{eqn-30}, the local LTL formula $\overbar{\varphi}_{i}$ is satisfied for $\mathbf{T}_{\tau}(\Sigma, \mathbbmss{P}_{i})$. 
\end{IEEEproof}

\begin{algorithm}[!t]
\DontPrintSemicolon\small
\caption{Local Abstract Controller (\texttt{LAC})}
\label{alg-4}
\KwIn{$\tau, \varepsilon\in\mathbb{R}^{+}, \mathbbmss{P}_{i}\in\mathbbmss{P}, \mathbf{O}\subset\mathbb{X}, \mathbbmss{R}(\pi_{\varepsilon})\subset\mathbb{X}$}
\KwOut{the local abstract controller $\mathscr{C}^{i}_{\mathsf{a}}$}
Construct the local symbolic abstraction $\mathcal{T}_{\tau, \mu_{i}, \eta_{i}}(\Sigma, \mathbbmss{P}_{i})$ to guarantee either the FRR or the $\varepsilon$-ABR \;
\eIf{$i=0$}{
Set $\mathbbmss{P}^{\varepsilon}_{0}\setminus\mathbf{O}$ as the local initial region \;
}{
Set $(\mathbbmss{P}^{\varepsilon}_{i-1}\cap\mathbbmss{P}^{\varepsilon}_{i})\setminus\mathbf{O}$ as the local initial region \;
}
\For{\emph{any element} $\mathbbmss{R}_{j}(\pi_{\varepsilon})$ \emph{in} $\mathbbmss{R}(\pi_{\varepsilon})$}{
\eIf{$\mathbbmss{P}^{\varepsilon}_{i}\cap\mathbbmss{R}_{j}(\pi_{\varepsilon})\neq\varnothing$}{
Set $(\mathbbmss{P}^{\varepsilon}_{i}\cap\mathbbmss{R}_{j}(\pi_{\varepsilon}))\setminus\mathbf{O}$ as the local internal regions \;
}{
No local internal regions \;
}
}
\eIf{$\mathbbmss{P}_{i+1}$ \emph{exists}}{
Set $(\mathbbmss{P}^{\varepsilon}_{i}\cap\mathbbmss{P}^{\varepsilon}_{i+1})\setminus\mathbf{O}$ as the local target region \;
}{
No local target region \;
}
Generate the local LTL formula $\overbar{\varphi}^{\varepsilon}_{i}$ based on $\pi_{\varepsilon}$  \;
Design the controller $\mathscr{C}^{i}_{\mathsf{a}}$ for $\mathcal{T}_{\tau, \mu_{i}, \eta_{i}}(\Sigma, \mathbbmss{P}^{\varepsilon}_{i})$ with $\overbar{\varphi}^{\varepsilon}_{i}$ \;
\textbf{return} the local abstract controller $\mathscr{C}^{i}_{\mathsf{a}}$
\end{algorithm}

From Theorems \ref{thm-4}-\ref{thm-5}, the essence of the controller synthesis is to design all local abstract controllers. Since each local symbolic abstraction has its own local LTL formula and all local LTL formulas are independent with each other, existing algorithms like the fixed-point algorithm \cite{Rungger2016scots} and the dynamic programming algorithm \cite{Bertsekas2005dynamic} can be applied locally to derive local abstract controllers. Therefore, the design of all local abstract controllers is addressed in the following.

For each cell $\mathbbmss{P}_{i}$, $i\in\mathbbmss{L}$, the local abstract controller design is presented in Algorithm \ref{alg-4}. First, the local symbolic abstraction $\mathcal{T}_{\tau, \mu_{i}, \eta_{i}}(\Sigma, \mathbbmss{P}_{i})$ is constructed in line 1 to guarantee certain equivalence relation (see also Theorems \ref{thm-2}-\ref{thm-3}). Second, the local LTL formula $\overbar{\varphi}^{\varepsilon}_{i}$ is formulated in lines 2-15 for $\mathcal{T}_{\tau, \mu_{i}, \eta_{i}}(\Sigma, \mathbbmss{P}_{i})$. In particular, lines 2-5 show the local initial region in two different cases, lines 6-10 determine all local internal regions in $\mathbbmss{P}_{i}$, and lines 11-14 provide the local target region in two different cases. All these regions have been explained in detail in Section \ref{subsec-LocalLTL}, and are used to construct the local LTL formula $\overbar{\varphi}^{\varepsilon}_{i}$ in line 15 based on the requirements of the accepting path $\pi_{\varepsilon}$. Finally, with the local symbolic model $\mathcal{T}_{\tau, \mu_{i}, \eta_{i}}(\Sigma, \mathbbmss{P}^{\varepsilon}_{i})$ and its local LTL formula $\overbar{\varphi}^{\varepsilon}_{i}$, the local abstract controller is designed in line 16 such that $\overbar{\varphi}^{\varepsilon}_{i}$ is satisfied for $\mathcal{T}_{\tau, \mu_{i}, \eta_{i}}(\Sigma, \mathbbmss{P}^{\varepsilon}_{i})$. Here, we emphasize that the local abstract controller is designed for $\mathcal{T}_{\tau, \mu_{i}, \eta_{i}}(\Sigma, \mathbbmss{P}^{\varepsilon}_{i})$ such that the abstract controller refinement in Theorem \ref{thm-5} can be implemented well.

Since any two successive cells in $\mathbbmss{P}$ overlap, the local target region in $\mathbbmss{P}_{i}$ will be the local initial region in $\mathbbmss{P}_{i+1}$, which is shown via lines 5 and 12 in Algorithm \ref{alg-4}. Therefore, Algorithm \ref{alg-4} can be implemented iteratively to derive all local abstract controllers. In this respect, the whole control strategy is summarized in Algorithm \ref{alg-5} to solve \textbf{Problems 1} and \textbf{2}. More precisely, if the path $\pmb{\mathbbmss{p}}$ from lines 1-4 does not exist, then Algorithm \ref{alg-5} terminates and returns a signal \textsf{Null} to show the unrealization of the accepting path for the LTL specification; otherwise, all local abstract controllers are designed in lines 7-10. Therefore, the following theorem is established directly.

\begin{algorithm}[!t]
\DontPrintSemicolon\small
\caption{Abstract Controller Design}
\label{alg-5}
\KwIn{$\tau, \varepsilon\in\mathbb{R}^{+}, \mathbb{X}_{0}\subseteq\mathbb{X}, \mathbbmss{R}(\pi_{\varepsilon})\subset\mathbb{X}, \mathbf{O}\subset\mathbb{X}$}
\KwOut{the union of all local abstract controllers}
\KwInit{$\mathscr{C}_{\mathsf{a}}=\mathscr{C}^{i}_{\mathsf{a}}=\varnothing$}
Cover the state space $\mathbb{X}$ via Algorithm \ref{alg-1} \;
Generate the graph $\mathcal{G}$ via Algorithm \ref{alg-3} \;
Generalize the graph $\mathcal{G}$ to $\bar{\mathcal{G}}$ by including $\mathbb{X}_{0}$ and $\mathbbmss{R}(\pi_{\varepsilon})$  \;
\eIf{\emph{the path $\pmb{\mathbbmss{p}}$ in Theorem \ref{thm-1} does not exist}}{
Stop and \textbf{return} \textsf{Null} \;
}{
Derive the cell sequence $\mathbbmss{P}$ from the path $\pmb{\mathbbmss{p}}$ \;
\For{\emph{all $\mathbbmss{P}_{i}\in\mathbbmss{P}$}}{
Design the local abstract controller $\mathscr{C}^{i}_{\mathsf{a}}=\texttt{LAC}(\tau, \mathbbmss{P}_{i}, \mathbf{O}, \mathbbmss{R}(\pi_{\varepsilon}))$ via Algorithm \ref{alg-4} \;
$\mathscr{C}_{\mathsf{a}}=\mathscr{C}_{\mathsf{a}}\cup\mathscr{C}^{i}_{\mathsf{a}}$ \;
}
\textbf{return} $\mathscr{C}_{\mathsf{a}}$
}
\end{algorithm}

\begin{theorem}
\label{thm-6}
Consider the system $\Sigma$, the obstacle set $\mathbb{O}\subset\mathbb{X}$, and the LTL formula $\varphi$ with the accepting path $\pi$. The realization of $\pi$ and the satisfaction of $\varphi$ can be verified from the result of Algorithm \ref{alg-5}. That is, Algorithm \ref{alg-5} returns either the signal \textsf{Null} to show the unrealization of $\pi$ or a sequence of local abstract controllers to guarantee the satisfaction of $\varphi$.
\end{theorem}

\begin{IEEEproof}
Follows by Theorems \ref{thm-1}, \ref{thm-4} and \ref{thm-5}.
\end{IEEEproof}

Theorem \ref{thm-6} shows that the satisfaction of the LTL formula $\varphi$ can be determined via Algorithm \ref{alg-5}, which thus solves \textbf{Problems 1} and \textbf{2}. In particular, all local abstract controllers can be combined as \eqref{eqn-27}, which can be refined as a global controller for the original system via \eqref{eqn-28}-\eqref{eqn-30}.

\section{Numerical Example}
\label{sec-example}

To illustrate the derived results, a numerical example is presented below. All computations are performed on a desk computer with Intel Core i9-10900K CPU@3.70GHz and 64GB RAM. Consider an autonomous vehicle whose dynamics is of the bicycle model in \cite[Example 2.8]{Astrom2010feedback}. To be specific, the system dynamics $\Sigma$ of the vehicle is given by
\begin{align}
\label{eqn-31}
\begin{bmatrix}\dot{x}_{1}\\ \dot{x}_{2}\\ \dot{x}_{3}\end{bmatrix}=\begin{bmatrix}
u_{1}\cos(\alpha+x_{3})\cos(\alpha)^{-1} \\
u_{1}\sin(\alpha+x_{3})\cos(\alpha)^{-1} \\
u_{1}\tan(u_{2})
\end{bmatrix},
\end{align}
where $\alpha:=\arctan(0.5\tan(u_{2}))$, $x:=(x_{1}, x_{2}, x_{3})\in\mathbb{R}^{3}$ is the vehicle state, and $u=(u_{1}, u_{2})\in\mathbb{R}^{2}$ is the control input. Here, $\mathbf{p}=(x_{1}, x_{2})\in\mathbb{R}^{2}$ is the vehicle position, $x_{3}\in\mathbb{R}$ is the orientation of the vehicle in the 2-dimensional plane, $u_{1}\in\mathbb{R}$ is the rear wheel velocity, and $u_{2}\in\mathbb{R}$ is the steering angle. The control input is assumed in $\mathbb{U}=[-1, 1]\times[-1, 1]$.

\begin{figure}[!t]
\begin{center}
\begin{picture}(110, 108)
\put(-100, -15){\resizebox{110mm}{45mm}{\includegraphics[width=2.5in]{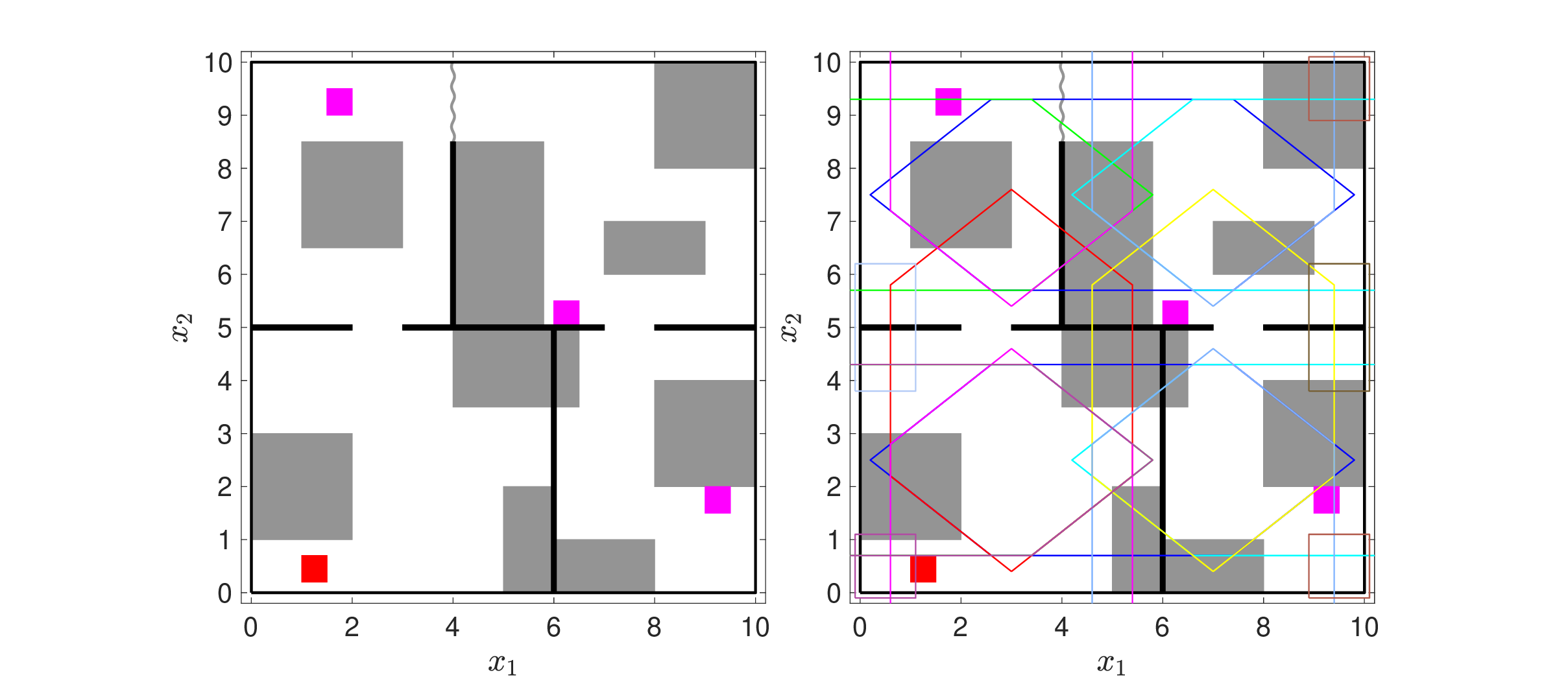}}}
\end{picture}
\end{center}
\caption{Illustration of the state space and its cover in the 2-D plane. (Left) The state space in the 2-D plane. The grey regions are obstacles, the red region is the initial region, the black lines are the walls among different rooms, and the grey wave line is the door. The magenta regions are the regions of interest. (Right) The covering of the state space in the 2-D plane.}
\label{fig-7}
\end{figure}

\begin{figure}[!t]
\begin{center}
\begin{picture}(55, 85)
\put(-55, -11){\resizebox{55mm}{36mm}{\includegraphics[width=2.5in]{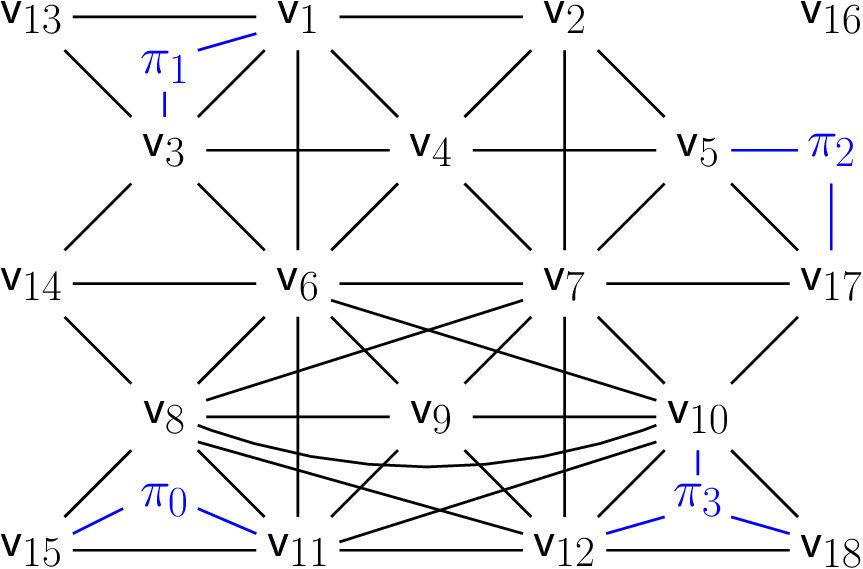}}}
\end{picture}
\end{center}
\caption{Illustration of the generated graph. $\mathcal{V}=\{\mathsf{v}_{i}: i=1, \ldots, 18\}$ is the vertex set. $\Pi=\{\pi_{0}, \pi_{1}, \pi_{2}, \pi_{3}\}$ is the symbol set of all regions of interest.}
\label{fig-8}
\end{figure}

\subsubsection{Problem Setup}
For the vehicle, the state space is $\mathbb{X}=[0, 10]\times[0, 10]\times[-\boldsymbol{\pi}, \boldsymbol{\pi}]$. In the 2-dimensional plane, the state space consists of 4 rooms, and there exists a door between Rooms 2 and 3 (i.e., the grey wave line in Fig. \ref{fig-7}). Initially, the vehicle is placed in the charge station $\mathbb{X}_{0}=[1, 1.5]\times[0.2, 0.7]\times[-\boldsymbol{\pi}, \boldsymbol{\pi}]$ of Room 1 (i.e., the red region in Fig. \ref{fig-7}) and the velocity is zero. The vehicle is to accomplish the following task: ``go to the region $\mathbf{S}_{1}=[1.5, 2]\times[9, 9.5]\times[-\boldsymbol{\pi}, \boldsymbol{\pi}]$ in Room 2, then visit the region $\mathbf{S}_{2}=[9.5, 10]\times[6, 6.5]\times[-\boldsymbol{\pi}, \boldsymbol{\pi}]$ in Room 3 or $\mathbf{S}_{3}=[8.5, 9]\times[0.5, 1]\times[-\boldsymbol{\pi}, \boldsymbol{\pi}]$ in Room 4, and finally stay in the region $\mathbf{S}_{3}$ while avoiding the obstacle regions in $\mathbb{O}$ (i.e., the black regions in Fig. \ref{fig-7})''. This task can be represented formally into the following LTL formula:
\begin{align}
\label{eqn-32}
\varphi&=(\neg(\pi_{2}\vee\pi_{3})\mathsf{U}\pi_{1})\wedge\diamondsuit(\pi_{2}\vee\pi_{3})\wedge\square\pi_{3},
\end{align}
where $\mathbbmss{R}(\pi_{1})=\mathbf{S}_{1}, \mathbbmss{R}(\pi_{2})=\mathbf{S}_{2}$ and $\mathbbmss{R}(\pi_{3})=\mathbf{S}_{3}$. For the LTL formula $\varphi$, we can implement \cite[Algorithm 3]{Guo2015multi} to derive an accepting path $\pi=\pi_{0}\pi_{1}\pi_{2}(\pi_{3})^{\omega}$, where $\mathbbmss{R}(\pi_{0})=\mathbb{X}_{0}$.

\subsubsection{Realization of the Accepting Path}
To illustrate the realization of the accepting path $\pi$, we choose $\epsilon=0.2$ and the state space is divided into 18 overlapping zonotopes, which is shown in Fig. \ref{fig-7}. Due to the existence of the door, we consider two cases to show the necessity of the realization verification.

The first case is that the door between Rooms 2 and 3 is open. In this case, we apply Algorithm \ref{alg-3} to generate the graph, which is shown in Fig. \ref{fig-8}. We search the graph in Fig. \ref{fig-8} to find a path to realize the LTL formula. We follow Theorem \ref{thm-1} and the shortest path search algorithm to find a path $\pmb{\mathbbmss{p}}:=\pi_{0}\mathsf{v}_{11}\mathsf{v}_{6}\mathsf{v}_{1}\pi_{1}\mathsf{v}_{2}\mathsf{v}_{5}\pi_{2}\mathsf{v}_{7}(\pi_{3})^{\omega}$. In particular, for all $\mathsf{v}_{k}\in\pmb{\mathbbmss{p}}$, $\mathbbmss{R}(\mathsf{v}_{k})\setminus\mathbb{O}\neq\varnothing$, which shows the validity of item (ii) of Theorem \ref{thm-1}. With the region $\mathbbmss{P}$ from the path $\pmb{\mathbbmss{p}}$, we follow Section \ref{subsec-LocalLTL} to construct all the local LTL formulas. For instance, in the region $\mathbbmss{R}(\mathsf{v}_{11})$, the local LTL formula is $\overbar{\varphi}_{1}=\overbar{\varphi}_{11}\wedge\overbar{\varphi}_{12}$ with $\overbar{\varphi}_{11}=\square(x\in\mathbbmss{R}(\mathsf{v}_{11}))$ and $\overbar{\varphi}_{12}=\diamondsuit(x\in\mathbbmss{R}(\mathsf{v}_{11})\cap\mathbbmss{R}(\mathsf{v}_{6}))$. Note that no local internal regions exist in $\mathbbmss{R}(\mathsf{v}_{11})$ and thus $\overbar{\varphi}_{13}$ does not exist. Similarly, we derive the local LTL formulas for all cells in $\mathbbmss{P}$, and combine them into $\overbar{\varphi}_{0}\wedge\overbar{\varphi}_{1}\wedge\ldots\wedge\overbar{\varphi}_{6}$. 

The second case is that the door is closed. From Fig. \ref{fig-8}, the path $\pmb{\mathbbmss{p}}$ in the first case is available to validate item (i) of Theorem \ref{thm-1}. However, the cell $\mathbbmss{R}(\mathsf{v}_{1})$ is not connected and no subregion in $\mathbbmss{R}(\mathsf{v}_{1})$ exists such that item (ii-b) of Theorem \ref{thm-1} is guaranteed. Hence, the accepting path $\pi$ is not realized and there is no need for the controller design in this case.

\begin{figure}[!t]
\begin{center}
\begin{picture}(70, 118)
\put(-70, -18){\resizebox{70mm}{50mm}{\includegraphics[width=2.5in]{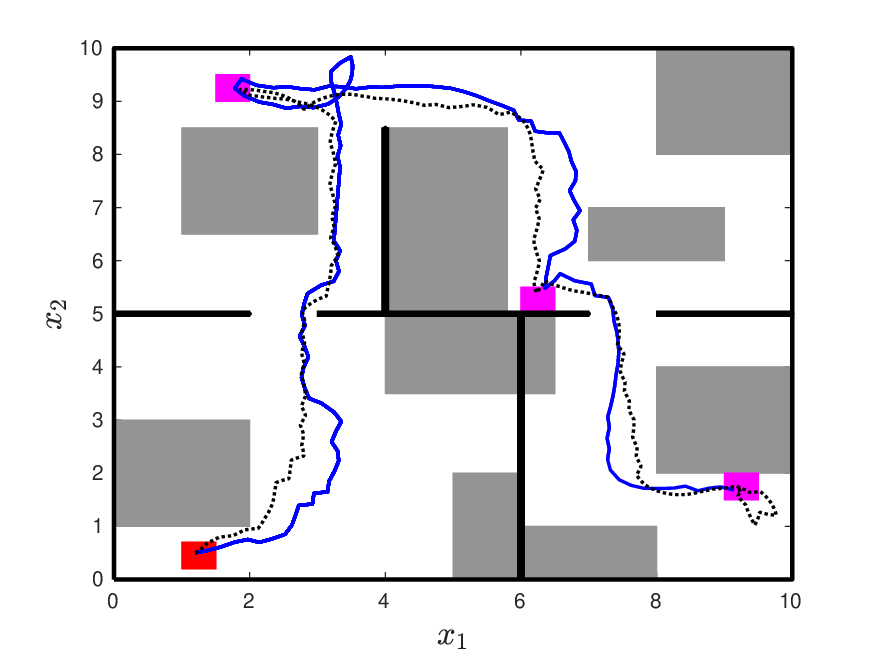}}}
\end{picture}
\end{center}
\caption{The state trajectories of the vehicle in the 2-D position space. The blue solid trajectory is generated via the proposed approach, while the black dotted one is based on the global approach.}
\label{fig-9}
\end{figure}

\begin{figure}[!t]
\begin{center}
\begin{picture}(80, 90)
\put(-70, -24){\resizebox{80mm}{45mm}{\includegraphics[width=2.5in]{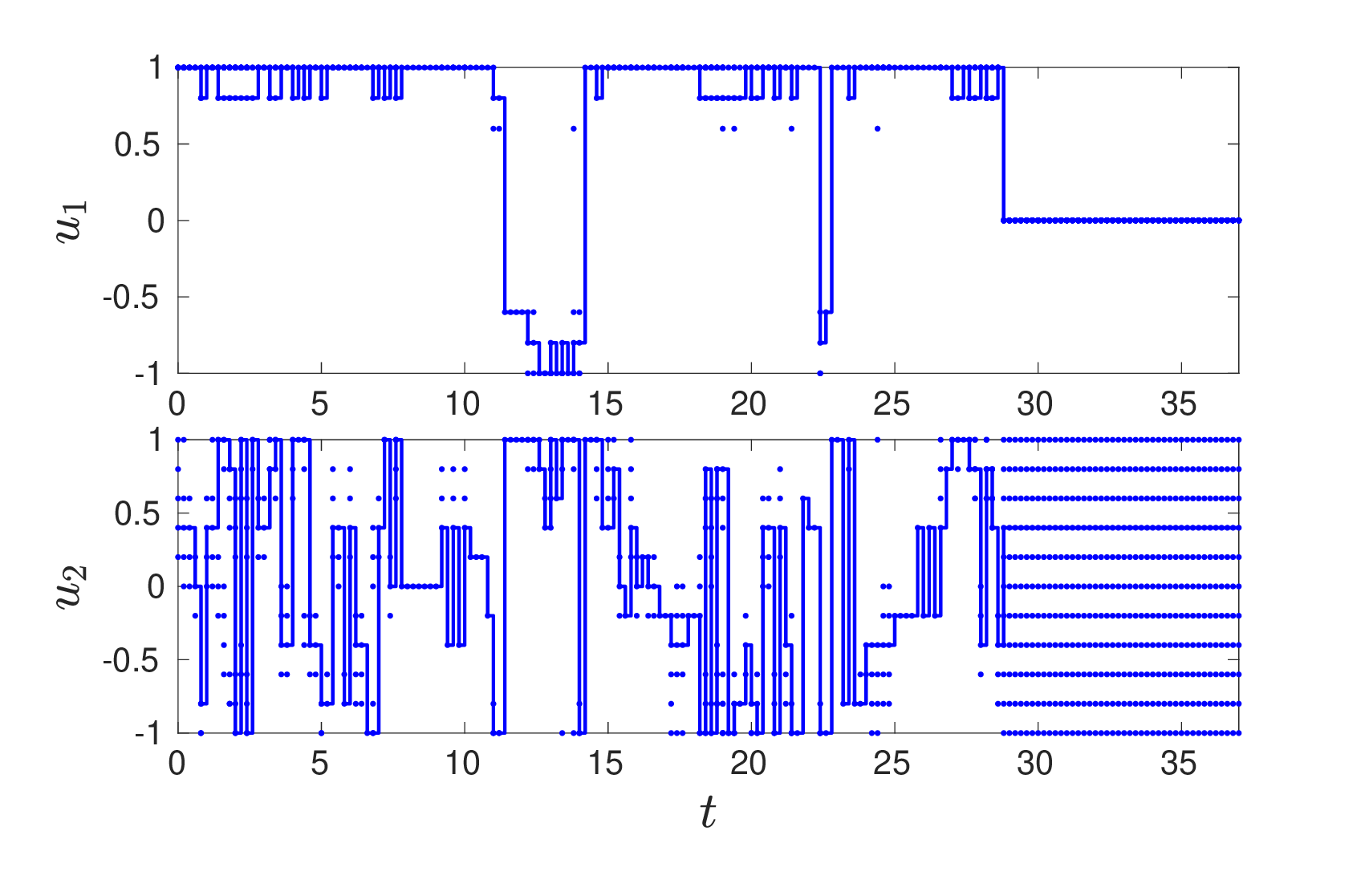}}}
\end{picture}
\end{center}
\caption{The control inputs for the satisfaction of the LTL formula $\varphi$. The blue dots are the feasible control inputs at each time instant, and the blue lines are the control trajectories to generate the blue position trajectory in Fig. \ref{fig-9}.}
\label{fig-10}
\end{figure}

From these two cases, we can see that the output of Algorithm \ref{alg-5} is \textsf{Null} in the second case. Hence, there is no need for the controller design, and the non-existence of the controller comes from the unrealization of $\pi$. Different from the classic abstraction-based approach \cite{Girard2012controller, Ren2020symbolic, Pola2008approximately, Reissig2017feedback} without the realization verification, the proposed approach reveals the real reason for the non-existence of the controller. Due to the realization verification, the costs of the abstraction construction and controller design can be avoided. From these two cases, we show the necessity of the realization verification.

\begin{table}[tp]
\centering
\caption[caption]{Comparison of transition numbers and run times}\vspace{-10pt}
\label{tab-1}
\begin{center}
\begin{tabular}{c|c|c|c}
\hline
& Transition numbers  & $\mathfrak{t}_{\mathsf{abs}}$ (s)  & $\mathfrak{t}_{\mathsf{con}}$ (s)   \\ \hline\hline 
$\mathbbmss{R}(\mathsf{v}_{11})$  &  $3.38112\cdot10^{7}$   &   102.30    &   44.13  \\ \hline
$\mathbbmss{R}(\mathsf{v}_{6})$  &  $5.34214\cdot10^{7}$    &  291.91  &     99.10\\ \hline
$\mathbbmss{R}(\mathsf{v}_{1})$ &  $2.18706\cdot10^{7}$      &    79.82  &     56.44  \\ \hline
$\mathbbmss{R}(\mathsf{v}_{2})$  &  $3.35502\cdot10^{7}$    &  107.57  &    49.14 \\ \hline
$\mathbbmss{R}(\mathsf{v}_{7})$ &  $3.57037\cdot10^{7}$      &   134.82 &   48.06   \\ \hline
$\mathbbmss{R}(\mathsf{v}_{10})$  &  $3.69582\cdot10^{7}$    &  134.18 &   54.00  \\ \hline
Global approach           &  $2.17682\cdot10^{8}$    &    1425.18   &   6015.38 \\ \hline
\end{tabular}
\end{center}
\end{table}

\subsubsection{Controller Design}
In the following, we only consider the first case and design the controller such that the LTL formula $\varphi$ is satisfied. Let $\tau=0.2$ and $\varepsilon=\epsilon=0.2$. In order to compare with the global approach, we follow the toolbox SCOTS \cite{Rungger2016scots} to approximate the state space and to construct the symbolic models $\mathbf{T}_{\tau, \mu_{i}, \eta_{i}}(\Sigma, \mathbbmss{P}_{i})$ for $i\in\{1, \ldots, 6\}$. Let $\eta_{i}=0.2$, and the input space $\mathbb{U}$ is approximated into $U_{2}:=\{(0.2j, 0.2k)\in\mathbb{R}^{2}: j, k\in\{-5, \ldots, 5\}\}$. Let $\mu_{1}=\mu_{2}=\mu_{4}=0.16, \mu_{3}=\mu_{5}=0.18$ and $\mu_{6}=0.15$. Hence, all local symbolic models are constructed and all local abstract controllers can be designed via Algorithm \ref{alg-4} to satisfy the local LTL formulas. Fig. \ref{fig-9} shows the position trajectories of the vehicle via the proposed and global approaches respectively, and thus implies that the global LTL formula is satisfied. In particular, $\mu=0.15$ for the global approach; otherwise, the LTL formula cannot be satisfied. The control inputs are given in Fig. \ref{fig-10}, and the vehicle stays in $\mathbf{S}_{3}$ since 28.8 seconds.

The comparison between the proposed approach and the global approach (e.g., \cite{Pola2008approximately, Reissig2017feedback}) is shown in Table \ref{tab-1}, where $\mathfrak{t}_{\mathsf{abs}}$ and $\mathfrak{t}_{\mathsf{con}}$ are respectively the computation times (in seconds) of the abstraction construction and control synthesis. We can see significant decreases in the computation times and transition numbers due to our proposed approach. Since each cell is a subset of the state space and different cells are allowed to have their own partition parameters, the number of transition relations in each cell is much smaller than that in the whole state space. Note that $\varphi$ in \eqref{eqn-32} is not co-safe, and thus we only need to construct a fine local symbolic model for $\mathbbmss{R}(\mathsf{v}_{10})$ instead of for the whole state space, which results in the decrease of the computation time for the controller synthesis. For the co-safe LTL formula $\varphi^{\prime}=(\neg(\varphi_{2}\vee\varphi_{3})\mathsf{U}\varphi_{1})\wedge\diamondsuit(\varphi_{2}\vee\varphi_{3})\wedge\diamondsuit\varphi_{3}$, the global approach can be applied with the relaxed parameter $\mu=\varepsilon$. In this case, the computation times are still much smaller since we only focus on six cells.

\section{Conclusion}
\label{sec-conclusion}

In this paper we studied the controller synthesis problem for nonlinear control systems with linear temporal logical specifications. A zonotope-based covering of the state space was proposed, which allows the intersection between neighbour cells and results in a graph among all cells. Further, a graph-based method was developed to verify the realization of the accepting paths for LTL specifications, which results in finite local LTL formulas via decomposition techniques. To ensure the satisfaction of LTL formulas, we discussed the relations between local and global controllers, and proposed a local-to-global control synthesis strategy. Future work will be devoted to the extension of the proposed approach to the case of multi-robot systems and the case where other temporal logic tasks like signal temporal logic are involved.

\appendices
\section{Proof of Theorem \ref{thm-1}}
\label{prf-thm1}

\setcounter{equation}{0}
\renewcommand{\theequation}{A.\arabic{equation}}

\begin{figure}[!t]
\begin{center}
\begin{picture}(60, 115)
\put(-62, -15){\resizebox{64mm}{44mm}{\includegraphics[width=2.5in]{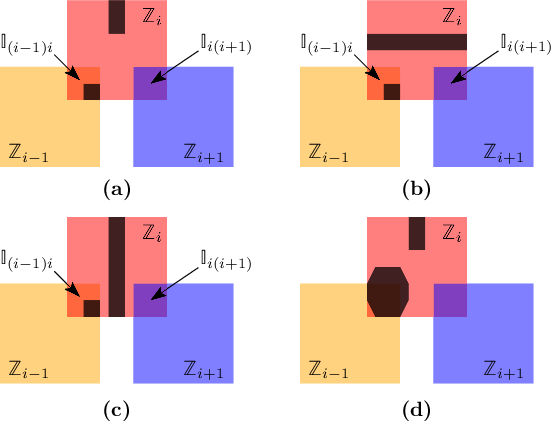}}}
\end{picture}
\end{center}
\caption{Illustration of different cases in the proof of Theorem \ref{thm-1}. The black regions are obstacles. \textbf{(a)} $\mathbbmss{P}_{i}\setminus\mathbf{O}$ is connected and item (ii-a) is satisfied. \textbf{(b)} $\mathbbmss{P}^{\prime}_{i}$ exists and item (ii-b) is satisfied. \textbf{(c)} $\mathbbmss{P}_{i}$ is not connected and $\mathbbmss{P}^{\prime}_{i}$ does not exist, which imply that item (ii-b) is not satisfied. \textbf{(d)} The intersection region $\mathbbmss{P}_{i}\cap\mathbbmss{P}_{i-1}$ is not admissible and the edge between $\mathbbmss{P}_{i}$ and $\mathbbmss{P}_{i-1}$ does not exist, which shows that item (i) is not satisfied.}
\label{fig-11}
\end{figure}

First, from Definition \ref{def-7}, the realization of the robust accepting path $\pi_{\varepsilon}$ implies there exists a connected region $\mathbbmss{X}\subset\mathbb{X}\setminus\mathbb{O}$ intersecting with all components in $\mathbbmss{R}(\pi_{\varepsilon})$. The existence of the connected region $\mathbbmss{X}$ implies the existence of the path $\pmb{\mathbbmss{p}}$ in $\bar{\mathcal{G}}$ such that $\pi_{\varepsilon}$ is realized, which shows the validity of item (i). To show item (ii), we consider two cases. The first case is that the region $\mathbbmss{P}$ from the path $\pmb{\mathbbmss{p}}$ is connected; see Fig. \ref{fig-11}(a). In this case, it readily follows that item (ii-a) holds. The second case is that the region $\mathbbmss{P}$ is not connected; see Fig. \ref{fig-11}(b). In this case, $\mathbbmss{X}\subset\mathbbmss{P}$, and for each $\mathbbmss{P}_{i}\in\mathbbmss{P}$, $\mathbbmss{X}\cap\mathbbmss{P}_{i}$ is connected, which implies the existence of $\mathbbmss{P}^{\prime}_{i}\subset\mathbbmss{P}^{\varepsilon}_{i}\setminus\mathbf{O}$ in item (ii-b) and further $\mathbbmss{P}^{\prime}_{i}\cup\mathbbmss{I}^{\varepsilon}_{(i-1)i}\cup\mathbbmss{I}^{\varepsilon}_{i(i+1)}$ is connected. Since $\mathbbmss{X}$ intersects with all components in $\mathbbmss{R}(\pi_{\varepsilon})$, we have $\mathbbmss{P}^{\prime}_{i}\cap\mathbbmss{R}(\pi_{\varepsilon})\neq\varnothing$ if $\mathbbmss{P}^{\varepsilon}_{i}\cap\mathbbmss{R}(\pi_{\varepsilon})\neq\varnothing$. Therefore, items (i)-(ii) are satisfied. In particular, if item (ii-b) does not hold, then the case in Fig. \ref{fig-11}(c) may exist, which contradicts with the realization of $\pi_{\varepsilon}$.

Second, we consider the realization of $\pi_{\varepsilon}$ from items (i)-(ii). If item (i) is satisfied, then for each pair $(\pi^{\varepsilon}_{k}, \pi^{\varepsilon}_{k+1})$ from the accepting path $\pi_{\varepsilon}$, there exists a sub-path in $\pmb{\mathbbmss{p}}$ such that $\pi^{\varepsilon}_{k}$ and $\pi^{\varepsilon}_{k+1}$ are connected, which in turn shows the case in Fig. \ref{fig-11}(d) is not feasible. In the following, we show the existence of the connected region $\mathbbmss{X}$ in the region $\mathbbmss{P}$ from the path $\pmb{\mathbbmss{p}}$. From item (ii-a), if $\mathbbmss{P}^{\varepsilon}_{i}\setminus\mathbf{O}$ is connected for each component $\mathbbmss{P}_{i}\in\mathbbmss{P}$ (see Fig. \ref{fig-11}(a)), then the region $\mathbbmss{P}$ is connected and we can set $\mathbbmss{X}=\mathbbmss{P}^{\varepsilon}\setminus\mathbf{O}$ directly. Otherwise, from item (ii-b), there exists a connected subregion $\mathbbmss{P}^{\prime}_{i}\subset\mathbbmss{P}_{i}\setminus\mathbf{O}$ intersecting with both $\mathbbmss{I}^{\varepsilon}_{(i-1)i}$ and $\mathbbmss{I}^{\varepsilon}_{i(i+1)}$; see Fig. \ref{fig-11}(b). In addition, if $\mathbbmss{P}^{\varepsilon}_{i}\cap\mathbbmss{R}(\pi_{\varepsilon})\neq\varnothing$, then $\mathbbmss{P}^{\prime}_{i}\cap\mathbbmss{R}(\pi_{\varepsilon})\neq\varnothing$. Hence, there exists a connected subregion in $\mathbbmss{P}\setminus\mathbf{O}$ intersecting with all components in $\mathbbmss{R}(\pi_{\varepsilon})$, which further implies the realization of $\pi_{\varepsilon}$.

\section{Proof of Theorem \ref{thm-2}}
\label{prf-thm2}

From the definitions of $\mathbf{T}_{\tau}(\Sigma, \mathbbmss{P}_{i})$ and $\mathbf{T}_{\tau, \mu_{i}, \eta_{i}}(\Sigma, \mathbbmss{P}_{i})$, one has $U_{2}(\mathbbmss{P}_{i})\subseteq U_{1}(\mathbbmss{P}_{i})$. Let $(x_{1}, q_{1})\in\mathscr{F}$ with $x_{1}\in\mathbbmss{P}_{i}$ and $q_{1}\in X_{2i}$, and $\|x_{1}-q_{1}\|_{\mathbf{G}}\leq\varepsilon$ holds from \eqref{eqn-23}. For each $u\in U_{2}(q_{1})$, we obtain $u\in U_{2}(q_{1})\subseteq U_{2}(\mathbbmss{P}_{i})\subseteq U_{1}(\mathbbmss{P}_{i})$, and $\Delta_{2i}(q_{1}, u)\neq\varnothing$ holds from the definition of $U_{2}(q_{1})$. If $\Delta_{1}(x_{1}, u)=\varnothing$, then $u\notin U_{1}(x_{1})$, which further implies $u\notin U_{1}$ and thus results in a contradiction. Hence, $\Delta_{1}(x_{1}, u)\neq\varnothing$ and $u\in U_{1}(x_{1})$. We conclude that $U_{2}(q_{1})\subseteq U_{1}(x_{1})$.

Let $q_{1}, q_{2}\in X_{2i}$ and $u\in U_{2}(q_{1})$. Since $(x_{1}, q_{1})\in\mathscr{F}$, $\|x_{1}-q_{1}\|_{\mathbf{G}}\leq\varepsilon$. If $\|\Delta_{1}(x_{1}, u)-q_{2}\|_{\mathbf{G}}\leq\varepsilon$, then there exists $x_{2}:=\mathbf{x}(\tau, x_{1}, u)\in\mathbbmss{P}_{i}$ such that $\|x_{2}-q_{2}\|_{\mathbf{G}}\leq\varepsilon$ from the approximation of $\mathbbmss{P}_{i}$. That is, $(x_{2}, q_{2})\in\mathscr{F}$. Since $f$ in \eqref{eqn-2} satisfies the local Lipschitz condition, one has $\|\mathbf{x}(\tau, x_{1}, u)-\mathbf{x}(\tau, q_{1}, u)\|_{\mathbf{G}}\leq e^{L_{i}\tau}\|x_{1}-q_{1}\|_{\mathbf{G}}\leq e^{L_{i}\tau}\varepsilon$ with the local Lipschitz constant $L_{i}>0$. Hence,
\begin{align*}
\|q_{2}-\mathbf{x}(\tau, q_{1}, u)\|_{\mathbf{G}}&\leq\|\mathbf{x}(\tau, x_{1}, u)-\mathbf{x}(\tau, q_{1}, u)\|_{\mathbf{G}} \\
&\quad +\|\mathbf{x}(\tau, x_{1}, u)-q_{2}\|_{\mathbf{G}} \\
&\leq e^{L_{i}\tau}\|x_{1}-q_{1}\|_{\mathbf{G}}+\varepsilon \\
&\leq(1+e^{L_{i}\tau})\varepsilon,
\end{align*}
which implies $q_{2}\in\Delta_{2i}(q_{1}, u)$ from the construction of $\mathbf{T}_{\tau, \mu_{i}, \eta_{i}}(\Sigma, \mathbbmss{P}_{i})$. Therefore, the proof is completed.

\section{Proof of Theorem \ref{thm-3}}
\label{prf-thm3}

\setcounter{equation}{0}
\renewcommand{\theequation}{C.\arabic{equation}}

We show that the relation $\mathscr{R}:=\{(x, q)\in X_{1i}\times X_{2i}: \|x-q\|_{\mathbf{G}}\leq\varepsilon\}$ is an $\varepsilon$-ABR between $\mathbf{T}_{\tau}(\Sigma, \mathbbmss{P}_{i})$ and $\overbar{\mathbf{T}}_{\tau, \mu_{i}, \eta_{i}}(\Sigma, \mathbbmss{P}_{i})$. The relation $\mathscr{R}$ implies that $\mathscr{R}(X_{1i})=X_{2i}$, and further $\mathbbmss{P}_{i}\subseteq\cup_{q\in X_{2i}}\{y\in\mathbb{R}^{n}: \|y-q\|_{\mathbf{G}}\leq0.5\mu_{i}\}$ from the geometrical consideration. For any $x_{1}\in\mathbbmss{P}_{i}$, there exists $q_{1}\in X_{2i}$ such that $\|x_{1}-q_{1}\|_{\mathbf{G}}\leq0.5\mu_{i}\leq\varepsilon$ from \eqref{eqn-25}.

Consider $(x_{1}, q_{1})\in\mathscr{R}$ with $x_{1}\in\mathbbmss{P}_{i}$ and $u_{1}\in U_{1}(\mathbbmss{P}_{i})$ such that $x_{2}=\Delta_{1}(x_{1}, u_{1})\in\mathbbmss{P}_{i}$. Let $\bar{x}:=\mathbf{x}(\tau, q_{1}, u_{1})\in\mathbbmss{P}_{i}$. If $\max_{l\in\pmb{\ell}}\{\|g_{l}\|\}\leq\mu_{i}$, then there exists $v\in\mathscr{A}(\mathbbmss{P}_{i})$ such that
\begin{align}
\label{eqn-C1}
\|\bar{x}-v\|_{\mathbf{G}}=\|\mathbf{x}(\tau, q_{1}, u_{1})-v\|_{\mathbf{G}}\leq0.5\mu_{i},
\end{align}
which implies that $\bar{x}\in\reach(\tau, q_{1})$ and $v\in\mathbbmss{S}_{\eta}(\tau, q_{1})$. Given any $u_{2}\in U_{2}(\mathbbmss{P}_{i})$ such that $u_{2}=\psi(v)$, there exists $w:=\mathbf{x}(\tau, q_{1}, u_{2})\in\mathbbmss{P}_{i}$ such that
\begin{align}
\label{eqn-C2}
\|w-v\|_{\mathbf{G}}=\|\mathbf{x}(\tau, q_{1}, u_{2})-v\|_{\mathbf{G}}\leq0.5\eta_{i}.
\end{align}
Since $\mathbbmss{P}_{i}\subseteq\cup_{q\in X_{2i}}\{y\in\mathbb{R}^{n}: \|y-q\|_{\mathbf{G}}\leq0.5\mu_{i}\}$, there exists $q_{2}\in X_{2}$ such that
\begin{align}
\label{eqn-C3}
\|q_{2}-w\|_{\mathbf{G}}=\|q_{2}-\mathbf{x}(\tau, q_{1}, u_{2})\|_{\mathbf{G}}\leq0.5\mu_{i}.
\end{align}

Let $q_{2}=\Delta_{2}(q_{1}, u_{2})$. Since the system $\Sigma$ satisfies \eqref{eqn-24}, we yield from \eqref{eqn-C1}-\eqref{eqn-C3} that
\begin{align*}
\|x_{2}-q_{2}\|_{\mathbf{G}}&\leq\|\Delta_{1}(x_{1}, u_{1})-\Delta_{2}(q_{1}, u_{2})\|_{\mathbf{G}} \\
&\leq\|\Delta_{1}(x_{1}, u_{1})-\bar{x}\|_{\mathbf{G}}+\|\bar{x}-\Delta_{2}(q_{1}, u_{2})\|_{\mathbf{G}} \\
&=\|\mathbf{x}(\tau, x_{1}, u_{1})-\mathbf{x}(\tau, q_{1}, u_{1})\|_{\mathbf{G}} \\
&\quad +\|\bar{x}-\Delta_{2}(q_{1}, u_{2})\|_{\mathbf{G}} \\
&\leq\beta(\|x_{1}-q_{1}\|_{\mathbf{G}}, \tau) \\
&\quad +\|\bar{x}-v+v-w+w-\Delta_{2}(q_{1}, u_{2})\|_{\mathbf{G}} \\
&\leq\beta(\varepsilon, \tau)+0.5\eta_{i}+\mu_{i}.
\end{align*}
Therefore, we conclude from \eqref{eqn-26} that $(x_{2}, q_{2})\in\mathscr{R}$.

Consider any $(x_{1}, q_{1})\in\mathscr{R}$ with $x_{1}\in\mathbbmss{P}_{i}$. Choose any $u_{2}\in U_{2}(\mathbbmss{P}_{i})$ such that $q_{2}=\Delta_{2}(q_{1}, u_{2})\in\mathbbmss{P}_{i}$. From the construction of $\mathbf{T}_{\tau, \mu_{i}, \eta_{i}}(\Sigma, \mathbbmss{P}_{i})$, $\|q_{2}-\mathbf{x}(t, q_{1}, u_{2})\|_{\mathbf{G}}\leq0.5\mu_{i}$. Pick $u_{1}=u_{2}$ and let $x_{2}=\Delta_{1}(x_{1}, u_{1})\in\mathbbmss{P}_{i}$. From \eqref{eqn-24}, we have
\begin{align*}
\|x_{2}-q_{2}\|_{\mathbf{G}}&=\|x_{2}-\mathbf{x}(t, q_{1}, u_{2})+\mathbf{x}(t, q_{1}, u_{2})-q_{2}\|_{\mathbf{G}}  \\
&\leq\|x_{2}-\mathbf{x}(t, q_{1}, u_{2})\|_{\mathbf{G}}+0.5\mu_{i} \\
&\leq\beta(\varepsilon, \tau)+0.5\mu_{i},
\end{align*}
which shows $(x_{2}, q_{2})\in\mathscr{R}$ from \eqref{eqn-26}, and completes the proof.



\begin{IEEEbiography}{Wei Ren}
received his B.Sci. degree from Hubei University, China, and his Ph.D. degree from the University of Science and Technology of China, China, in 2011 and 2018. He was a visiting student at the University of Melbourne, Australia, from 2017 to 2018. From 2018 to 2022, he was a postdoctoral fellow at KTH Royal Institute of Technology, Stockholm, Sweden and Universit\'e Catholique de Louvain, Belgium. He currently is a Professor at Dalian University of Technology, Dalian, China. His research interests include Multi-Agent Systems, Networked Control, Formal Methods, and Hybrid Systems.
\end{IEEEbiography}

\begin{IEEEbiography}{Rapha\"el M. Jungers}
is an FNRS Professor at UCLouvain, Belgium. His main interests lie in the fields of Computer Science, Graph Theory, Optimization and Control. He received a Ph.D. in Mathematical Engineering from UCLouvain (2008), and a M.Sc. in Applied Mathematics, both from the EcoleCentrale Paris, (2004), and from UCLouvain (2005). He has held various invited positions, at the Universit\'e Libre de Bruxelles (2008-2009), at the Laboratory for Information and Decision Systems of the Massachusetts Institute of Technology (2009-2010), at the University of L'Aquila (2011, 2013, 2016), at the University of California Los Angeles (2016-2017), and at the University of Oxford (2022-2023). He is an FNRS, BAEF, and Fulbright fellow. He has been an Editor at large for  IEEE CDC, Associate Editor for the IEEE CSS Conference Editorial Board, and the journals NAHS, Systems and Control Letters, and IEEE Transactions on Automatic Control. He was the recipient of the IBM Belgium 2009 award and a finalist of the ERCIM Cor Baaten award 2011. He was the co-recipient of the SICON best paper award 2013-2014.
\end{IEEEbiography}

\begin{IEEEbiography}{Dimos V. Dimarogonas}
was born in Athens, Greece, in 1978. He received the Diploma in Electrical and Computer Engineering in 2001 and the Ph.D. in Mechanical Engineering in 2007, both from National Technical University of Athens (NTUA), Greece. Between 2007 and 2010, he held postdoctoral positions at the KTH Royal Institute, Stockholm Sweden, and the Laboratory for Information and Decision Systems (LIDS), MIT, Boston USA. He is currently a Professor and Head at the Division of Decision and Control, KTH Royal Institute of Technology, Stockholm, Sweden. His current research interests include Multi-Agent Systems, Hybrid Systems and Control, Robot Navigation and Networked Control. He serves in the Editorial Board of Automatica, and the IEEE Transactions on Control of Network Systems. He is an IEEE Fellow. 
\end{IEEEbiography}


\begin{thebibliography}{10}
\bibitem{Baier2008principles}
C.~Baier and J.-P. Katoen, \emph{Principles of Model Checking}. MIT press, 2008.

\bibitem{Wolff2014optimization}
E.~M. Wolff, U.~Topcu, and R.~M. Murray, ``Optimization-based trajectory generation with linear temporal logic specifications,'' in \emph{IEEE Int. Conf. Robot. Autom.} IEEE, 2014, pp. 5319--5325.

\bibitem{Bloem2012synthesis}
R.~Bloem, B.~Jobstmann, N.~Piterman, A.~Pnueli, and Y.~Sa'ar, ``Synthesis of reactive (1) designs,'' \emph{Journal of Computer and System Sciences}, vol.~78, no.~3, pp. 911--938, 2012.

\bibitem{Clarke2018handbook}
E.~M. Clarke, T.~A. Henzinger, H.~Veith, R.~Bloem \emph{et~al.}, \emph{Handbook of Model Checking}. Springer, 2018, vol.~10.

\bibitem{Wongpiromsarn2011tulip}
T.~Wongpiromsarn, U.~Topcu, N.~Ozay, H.~Xu, and R.~M. Murray, ``{TuLiP: A} software toolbox for receding horizon temporal logic planning,'' in \emph{Int. Conf. Hybrid Syst.: Comput. Control}, 2011, pp. 313--314.

\bibitem{Antsaklis1993hybrid}
P.~J. Antsaklis, M.~Lemmon, and J.~A. Stiver, ``Hybrid system modeling and event identification,'' \emph{Technical Report of the ISIS Group at the University of Notre Dame ISIS-93-002}, vol.~93, pp. 366--392, 1993.

\bibitem{Tabuada2009verification}
P.~Tabuada, \emph{Verification and Control of Hybrid Systems: A Symbolic Approach}.\hskip 1em plus 0.5em minus 0.4em\relax Springer Science \& Business Media, 2009.

\bibitem{Girard2012controller}
A.~Girard, ``Controller synthesis for safety and reachability via approximate bisimulation,'' \emph{Automatica}, vol.~48, no.~5, pp. 947--953, 2012.

\bibitem{Pola2008approximately}
G.~Pola, A.~Girard, and P.~Tabuada, ``Approximately bisimilar symbolic models for nonlinear control systems,'' \emph{Automatica}, vol.~44, no.~10, pp. 2508--2516, 2008.

\bibitem{Zamani2012symbolic}
M.~Zamani, G.~Pola, M.~Mazo, and P.~Tabuada, ``Symbolic models for nonlinear control systems without stability assumptions,'' \emph{IEEE Trans. Automat. Contr.}, vol.~57, no.~7, pp. 1804--1809, 2012.

\bibitem{Ren2019logarithmic}
W.~Ren and D.~V. Dimarogonas, ``Logarithmic quantization based symbolic abstractions for nonlinear control systems,'' in \emph{European Control Conference}. IEEE, 2019, pp. 1312--1317.

\bibitem{Pola2010symbolic}
G.~Pola, P.~Pepe, M.~D. Di~Benedetto, and P.~Tabuada, ``Symbolic models for nonlinear time-delay systems using approximate bisimulations,'' \emph{Systems \& Control Letters}, vol.~59, no.~6, pp. 365--373, 2010.

\bibitem{Ren2020symbolic}
W.~Ren and D.~V. Dimarogonas, ``Symbolic abstractions for nonlinear control systems via feedback refinement relation,'' \emph{Automatica}, vol. 114, p. 108828, 2020.

\bibitem{Girard2016safety}
A.~Girard, G.~G{\"o}ssler, and S.~Mouelhi, ``Safety controller synthesis for incrementally stable switched systems using multiscale symbolic models,'' \emph{IEEE Trans. Automat. Contr.}, vol.~61, no.~6, pp. 1537--1549, 2016.

\bibitem{Zamani2017symbolic}
M.~Zamani, M.~Mazo~Jr, M.~Khaled, and A.~Abate, ``Symbolic abstractions of networked control systems,'' \emph{IEEE Trans. Control. Netw. Syst.}, vol.~5, no.~4, pp. 1622--1634, 2017.

\bibitem{Zamani2015symbolic}
M.~Zamani, A.~Abate, and A.~Girard, ``Symbolic models for stochastic switched systems: {A} discretization and a discretization-free approach,'' \emph{Automatica}, vol.~55, pp. 183--196, 2015.

\bibitem{Reissig2017feedback}
G.~Reissig, A.~Weber, and M.~Rungger, ``Feedback refinement relations for the synthesis of symbolic controllers,'' \emph{IEEE Trans. Automat. Contr.}, vol.~62, no.~4, pp. 1781--1796, 2017.

\bibitem{Kim2017symbolic}
E.~S. Kim, M.~Arcak, and S.~A. Seshia, ``Symbolic control design for monotone systems with directed specifications,'' \emph{Automatica}, vol.~83, pp. 10--19, 2017.

\bibitem{Mazo2010pessoa}
M.~Mazo, A.~Davitian, and P.~Tabuada, ``Pessoa: {A} tool for embedded controller synthesis,'' in \emph{Int. Conf. Comput. Aided Verification}. Springer, 2010, pp. 566--569.

\bibitem{Mouelhi2013cosyma}
S.~Mouelhi, A.~Girard, and G.~G{\"o}ssler, ``{CoSyMA: A} tool for controller synthesis using multi-scale abstractions,'' in \emph{Int. Conf. Hybrid Syst.: Comput. Control}, 2013, pp. 83--88.

\bibitem{Rungger2016scots}
M.~Rungger and M.~Zamani, ``{SCOTS}: {A} tool for the synthesis of symbolic   controllers,'' in \emph{Int. Conf. Hybrid Syst.: Comput. Control}. ACM, 2016, pp. 99--104.

\bibitem{Tumova2010symbolic}
J.~Tumova, B.~Yordanov, C.~Belta, I.~{\v{C}}ern{\'a}, and J.~Barnat, ``A symbolic approach to controlling piecewise affine systems,'' in \emph{Proc. IEEE Conf. Decis. Control.} IEEE, 2010, pp. 4230--4235.

\bibitem{Scott2016constrained}
J.~K. Scott, D.~M. Raimondo, G.~R. Marseglia, and R.~D. Braatz, ``Constrained zonotopes: {A} new tool for set-based estimation and fault detection,'' \emph{Automatica}, vol.~69, pp. 126--136, 2016.

\bibitem{Girard2005reachability}
A.~Girard, ``Reachability of uncertain linear systems using zonotopes,'' in \emph{Int. Conf. Hybrid Syst.: Comput. Control}. Springer, 2005, pp. 291--305.

\bibitem{Mitchell2019invariant}
I.~M. Mitchell, J.~Budzis, and A.~Bolyachevets, ``Invariant, viability and discriminating kernel under-approximation via zonotope scaling,'' in \emph{Int. Conf. Hybrid Syst., Comput. Control}, 2019, pp. 268--269.

\bibitem{Ren2021zonotope}
W.~Ren, J.~Calbert, and R.~Jungers, ``Zonotope-based controller synthesis for {LTL} specifications,'' in \emph{IEEE Conf. Decis. Control}. IEEE, 2021, pp. 580--585.

\bibitem{Belta2007symbolic}
C.~Belta, A.~Bicchi, M.~Egerstedt, E.~Frazzoli, E.~Klavins, and G.~J. Pappas, ``Symbolic planning and control of robot motion,'' \emph{IEEE Robot. Autom. Mag.}, vol.~14, no.~1, pp. 61--70, 2007.

\bibitem{Kress2009temporal}
H.~Kress-Gazit, G.~E. Fainekos, and G.~J. Pappas, ``Temporal-logic-based reactive mission and motion planning,'' \emph{IEEE Trans. Robot.}, vol.~25, no.~6, pp. 1370--1381, 2009.

\bibitem{Fainekos2009temporal}
G.~E. Fainekos, A.~Girard, H.~Kress-Gazit, and G.~J. Pappas, ``Temporal logic motion planning for dynamic robots,'' \emph{Automatica}, vol.~45, no.~2, pp. 343--352, 2009.

\bibitem{Gol2013time}
E.~A. Gol and C.~Belta, ``Time-constrained temporal logic control of multi-affine systems,'' \emph{Nonlinear Anal. Hybrid Syst.}, vol.~10, pp. 21--33, 2013.

\bibitem{Meyer2019hierarchical}
P.-J. Meyer and D.~V. Dimarogonas, ``Hierarchical decomposition of {LTL} synthesis problem for nonlinear control systems,'' \emph{IEEE Trans. Automat. Contr.}, vol.~64, no.~11, pp. 4676--4683, 2019.

\bibitem{Girard2007approximation}
A.~Girard and G.~J. Pappas, ``Approximation metrics for discrete and continuous systems,'' \emph{IEEE Trans. Automat. Contr.}, vol.~5, no.~52, pp. 782--798, 2007.

\bibitem{Tabuada2008approximate}
P.~Tabuada, ``An approximate simulation approach to symbolic control,'' \emph{IEEE Trans. Automat. Contr.}, vol.~53, no.~6, pp. 1406--1418, 2008. 

\bibitem{Girard2010approximately}
A.~Girard, G.~Pola, and P.~Tabuada, ``Approximately bisimilar symbolic models for incrementally stable switched systems,'' \emph{IEEE Trans. Automat. Contr.}, vol.~55, no.~1, pp. 116--126, 2010.

\bibitem{Guo2015multi}
M.~Guo and D.~V. Dimarogonas, ``Multi-agent plan reconfiguration under local {LTL} specifications,'' \emph{Int. J. Robot. Res.}, vol.~34, no.~2, pp. 218--235, 2015.

\bibitem{Hsu2018lazy}
K.~Hsu, R.~Majumdar, K.~Mallik, and A.-K. Schmuck, ``Lazy abstraction-based control for safety specifications,'' in \emph{IEEE Conf. Decis. Control}. IEEE, 2018, pp. 4902--4907.

\bibitem{Lindemann2019coupled}
L.~Lindemann, J.~Nowak, L.~Sch{\"o}nb{\"a}chler, M.~Guo, J.~Tumova, and D.~V. Dimarogonas, ``Coupled multi-robot systems under linear temporal logic and signal temporal logic tasks,'' \emph{IEEE Trans. Control Syst. Technol.}, vol.~29, no.~2, pp. 858--865, 2019.

\bibitem{Rego2020guaranteed}
B.~S. Rego, G.~V. Raffo, J.~K. Scott, and D.~M. Raimondo, ``Guaranteed methods based on constrained zonotopes for set-valued state estimation of nonlinear discrete-time systems,'' \emph{Automatica}, vol. 111, p. 108614, 2020.

\bibitem{Kochdumper2019representation}
N.~Kochdumper and M.~Althoff, ``Representation of polytopes as polynomial zonotopes,'' \emph{arXiv preprint arXiv:1910.07271}, 2019.

\bibitem{Ren2019dynamic}
W.~Ren and D.~V. Dimarogonas, ``Dynamic quantization based symbolic abstractions for nonlinear control systems,'' in \emph{IEEE Conf. Decis. Control}. IEEE, 2019, pp. 4343--4348.

\bibitem{Mcmullen1975space}
P.~McMullen, ``Space tiling zonotopes,'' \emph{Mathematika}, vol.~22, no.~2, pp. 202--211, 1975.

\bibitem{Kabi2020synthesizing}
B.~Kabi, ``Synthesizing invariants: A constraint programming approach based on zonotopic abstraction,'' Ph.D. dissertation, Institut Polytechnique de Paris, 2020.

\bibitem{Richter1993line}
J.~Richter-Gebert, ``Line arrangements and zonotopal tilings: {A} little printer exercise,'' \emph{Hyper-Space}, vol.~2, pp. 8--17, 1993.

\bibitem{Bertsekas2005dynamic}
D.~P. Bertsekas, \emph{Dynamic Programming and Optimal Control}. Athena scientific Belmont, MA, 2005, vol.~1, no.~3.

\bibitem{Astrom2010feedback}
K.~J. Astr{\"o}m and R.~M. Murray, \emph{Feedback Systems: An Introduction for Scientists and Engineers}. Princeton University Press, 2010.
\end{thebibliography}
\end{document}